\documentclass{article}

\usepackage{PRIMEarxiv}

\usepackage[utf8]{inputenc} 
\usepackage[T1]{fontenc}    
\usepackage{hyperref}       
\usepackage{url}            
\usepackage{booktabs}       
\usepackage{amsfonts}       
\usepackage{nicefrac}       
\usepackage{microtype}      
\usepackage{lipsum}
\usepackage{fancyhdr}       
\usepackage{graphicx}       
\usepackage{subcaption}
\graphicspath{{media/}}     

\title{Image-to-Text Translation for Interactive Image Recognition: A Comparative User Study \\with Non-Expert Users}

\author{
  Wataru Kawabe and Yusuke Sugano \\
  Institute of Industrial Science \\
  The University of Tokyo \\
  Tokyo, Japan\\
  \texttt{\{wkawabe, sugano\}@iis.u-tokyo.ac.jp} \\
}

\begin{document}
\maketitle

\begin{abstract}
Interactive machine learning (IML) allows users to build their custom machine learning models without expert knowledge.
While most existing IML systems are designed with classification algorithms, they sometimes oversimplify the capabilities of machine learning algorithms and restrict the user's task definition.
On the other hand, as recent large-scale language models have shown, natural language representation has the potential to enable more flexible and generic task descriptions.
Models that take images as input and output text have the potential to represent a variety of tasks by providing appropriate text labels for training.
However, the effect of introducing text labels to IML system design has never been investigated.
In this work, we aim to investigate the difference between image-to-text translation and image classification for IML systems.
Using our prototype systems, we conducted a comparative user study with non-expert users, where participants solved various tasks.
Our results demonstrate the underlying difficulty for users in properly defining image recognition tasks while highlighting the potential and challenges of interactive image-to-text translation systems.
\end{abstract}

\keywords{Interactive Machine Learning \and Graphical User Interface}

\section{Introduction}

Recent advances in machine learning (ML) and deep neural networks have greatly expanded the opportunities for various real-world applications.
Despite this trend, providing an all-purpose ML model is still a challenging task, and it is important to provide a capability for diverse users to train user- and task-specific models.
As a potential solution to this challenge, interactive machine learning (IML) systems aim to provide a way for non-expert users without ML knowledge to interact with ML algorithms and to prototype their own ML models~\cite{fails2003interactive, amershi2014power}.
Many research efforts have been made to create interactive image, sound, and text recognition systems, and their effectiveness has been verified through extensive user studies~\cite{dudley2018review}. 

In most previous work, IML systems are designed based on classification algorithms~\cite{fiebrink2011human, arendt2019towards, kulesza2015principles, talbot2009ensemblematrix, kapoor2010interactive}.
Classification can be seen as the most straightforward ML formulation, and users can define target category labels and add their corresponding training samples.
However, the classification-based design has some fundamental limitations.
First, despite the simplicity, it has been noted that novice users may have difficulty defining abstract category labels, i.e., categories containing multiple categories of concrete objects, using a classification-based GUI~\cite{nakao2020use}.
For example, if the user wants to create an emergency vehicle detector, s/he needs to annotate binary labels indicating if the image contains an emergency vehicle or not.
However, novice users tend to assign annotations of concrete car categories, such as {\em ambulances} and {\em fire trucks}, instead of the abstract label {\em emergency vehicle}.
Second, fundamentally speaking, classification-based design strictly limits user-definable recognition tasks.
Especially when it comes to complex media, such as images, there are many recognition tasks that cannot be formulated as a classification.
This problem is not limited to classification-based systems; an interface specialized for one particular formulation may prevent the user from freely formulating recognition tasks.

The output format is one of the most important factors for defining ML tasks.
For example, in the case of image input, the model is expected to output bounding boxes and their corresponding object categories for object detection~\cite{zou2019object, zhao2019object, papageorgiou1998general}, while the typical output format for semantic segmentation is pixel-level classification maps~\cite{long2015fully, yu2018methods, taghanaki2021deep}.
As recent examples of language modeling studies have shown, text output has the advantage of being able to handle a wide variety of tasks in a generic manner.
It has been pointed out that large-scale language models are capable of handling a variety of tasks through in-context learning and fine-tuning~\cite{radford2019language, brown2020language}, and this is also true for cross-modal cases such as vision-language models~\cite{gan2022vision, lu2019vilbert, li2022blip}.
Image-to-text translation, or image captioning, has been actively studied in the computer vision community as one of the most fundamental image recognition tasks~\cite{hossain2019comprehensive, vinyals2015show, pan2004automatic}.
Although the phrase ``image captioning'' often refers to the specific task of describing the content of an image, the image-to-text translation model, which converts arbitrary images to text, can express a variety of tasks that go beyond the generic description of the content of the image.
For example, the model can be seen as a face detector if it outputs simple texts indicating the location of the faces in the input image.
Likewise, when a text describes the shapes of image regions together with the categories they belong to, it can be seen as a rough representation of a semantic segmentation map.
Compared to image classification, image-to-text translation backends can potentially extend the capability of interactive image recognition systems.
However, there is no prior work focusing on comparing these two algorithms from the IML perspective or designing an IML system based on image-to-text translation.

\begin{figure*}[t]
  \centering
  \includegraphics[width=\linewidth]{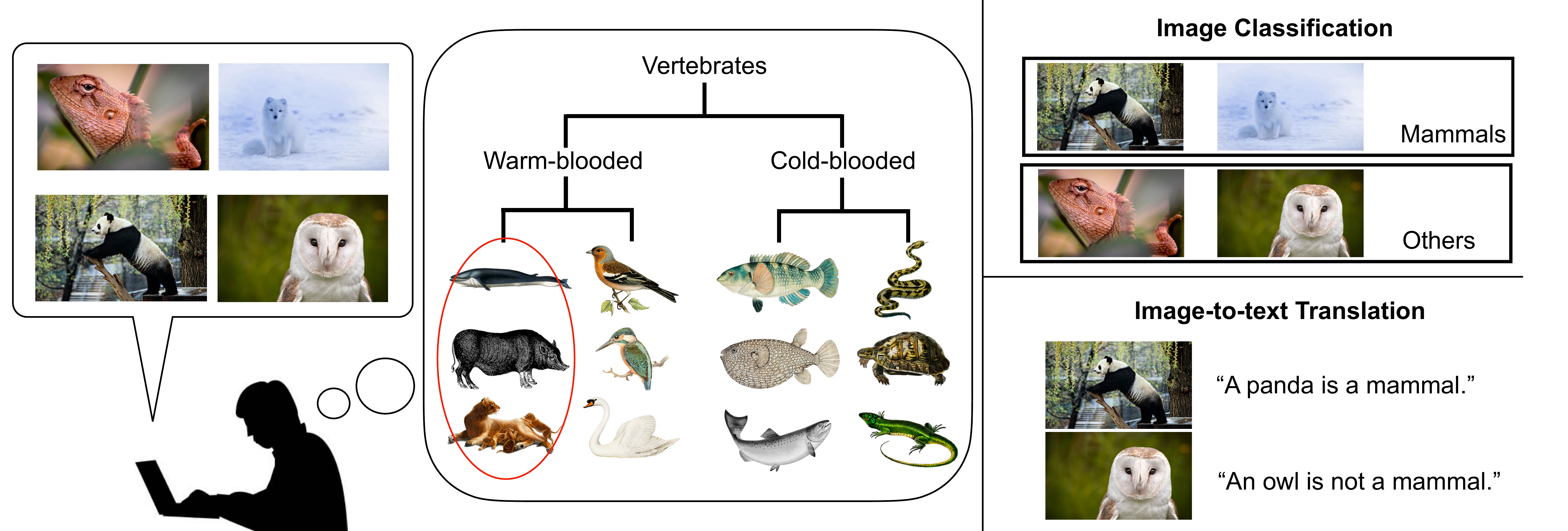}
  \caption{The goal of this work is to investigate a design of interactive image recognition systems based on text output. Using a novel interactive image-to-translation framework, we analyze whether such a design can address the limitations of classification-based systems.}
  \label{fig:teaser}
\end{figure*}

The purpose of this work is to examine how the user annotations and the system usability change when a model that produces text output is adopted in the design of an IML system instead of classification.
As shown in Fig.~\ref{fig:teaser}, this study compares two IML systems for image recognition. 
One is an \emph{image classification} system that provides category output, and the other is an \emph{image-to-text translation} system that provides text output.
We conducted a comparative user study using both image classification- and image-to-text translation-based prototype IML systems.
With both systems, we asked novice participants to freely train models to solve different types of image recognition tasks.
By analyzing the user annotation and training results, we confirmed that there are certainly some tasks that non-expert users have difficulty describing correctly.
Although this difficulty cannot be fully resolved by introducing text output, we also observed that users tend to provide richer information, which partly improves the task description process.
We also found that text output did not significantly reduce the usability of the interactive image recognition system, and users were able to perform annotations aligned with the system characteristics.
Throughout the analysis, we summarize the potential and challenges of using text as an output format in IML systems.

\section{Related Work}

\subsection{Interactive Machine Learning}
IML systems aim at getting users involved in the training process of ML models~\cite{dudley2018review, fails2003interactive, amershi2014power, ware2001interactive}.
Their typical objective is to let users define the recognition task and provide their training data.
As discussed above, many prior works employed classification-based design to provide simple and intuitive GUIs for image and/or sound recognition~\cite{talbot2009ensemblematrix, carney2020teachable, tatsuya2020investigating}.
While these works provide flexibility in the definition of the target to users, the flexibility of task description is inevitably limited to classification.
Some prior work proposed other forms of user-customizable IML systems where users can register their own target objects~\cite{kacorri2017people, ahmetovic2020recog, liu2022interactive}, create their own rules for image search~\cite{fogarty2008cueflik}, or customize feature space for data sorting~\cite{pirrung2018sharkzor, hodas2016adding}.
However, these cases still focus on task-specific customization scenarios and users cannot fully control the task definition.
The focus of this work is to explore the capability of IML systems for non-expert users to freely define their own recognition tasks.
To this end, we compare image classification- and image-to-text translation-based IML prototypes under the condition that participants solve various image recognition tasks.

Some prior works have also conducted user analysis to understand and improve IML systems~\cite{amershi2009overview, fiebrink2011human, kulesza2009fixing, patel2008investigating, nakao2020use}.
Most of these previous research asks participants/users to solve a single pre-defined task~\cite{kulesza2009fixing, patel2008investigating, fiebrink2011human}, or just let them design their own tasks with the system~\cite{nakao2020use, amershi2009overview}.
One unique aspect of this work is to ask participants to solve multiple diverse image recognition tasks with interactive systems and discuss the task-independent properties of system design.
In addition, this is the first work to report a detailed user study of the interactive image-to-text translation system.

\subsection{Image-to-Text Translation}
The description of images in natural language is one of the greatest challenges of image recognition, and image-to-text translation algorithms have been actively studied in the computer vision community~\cite{li2011composing, vinyals2015show, aneja2018convolutional, you2016image, cornia2020meshed, liu2020interactive}.
Although some previous work addressed the task of customizing image captioning models based on user input~\cite{cornia2019show, jia2020icap, hossain2021text}, their goal is still uniform scene descriptions, and users cannot fully control the output text for a wide range of descriptive contents in an image.
Cornia et al. proposed a captioning model that adaptively describes image content based on user-defined object bounding boxes~\cite{cornia2019show}.
Jia et al. proposed a human-in-the-loop image captioning system called iCap~\cite{jia2020icap}, which takes both the target image and an incomplete user description to predict complete captions.
Another closely related research direction is the adaptation of image captioning models to a different style of captions.
Some examples tried to achieve cross-domain image captioning, which transfers caption styles between unpaired data~\cite{chen2017show, zhao2020cross, long2020cross, yang2018multitask, zhao2017dual}.
There have also been some research examples focusing on fine-tuning pre-trained image captioning models to adapt to the small target training data~\cite{chen2021visualgpt, lu2019vilbert, li2020oscar}.
The ML backend of our interactive image-to-text translation system is similar to such fine-tuning models.
However, this work differs from these previous studies in that we design a GUI that allows users to freely describe and define the output text and conduct experiments to compare user behavior.
Unlike these prior works focusing on technical aspects, our study also incorporates a user study to analyze how such a model can be applied to the interactive image recognition scenario.

\section{Design of Interactive Image Recognition Systems}

\begin{figure*}[t]
  \centering
  \includegraphics[width=\linewidth]{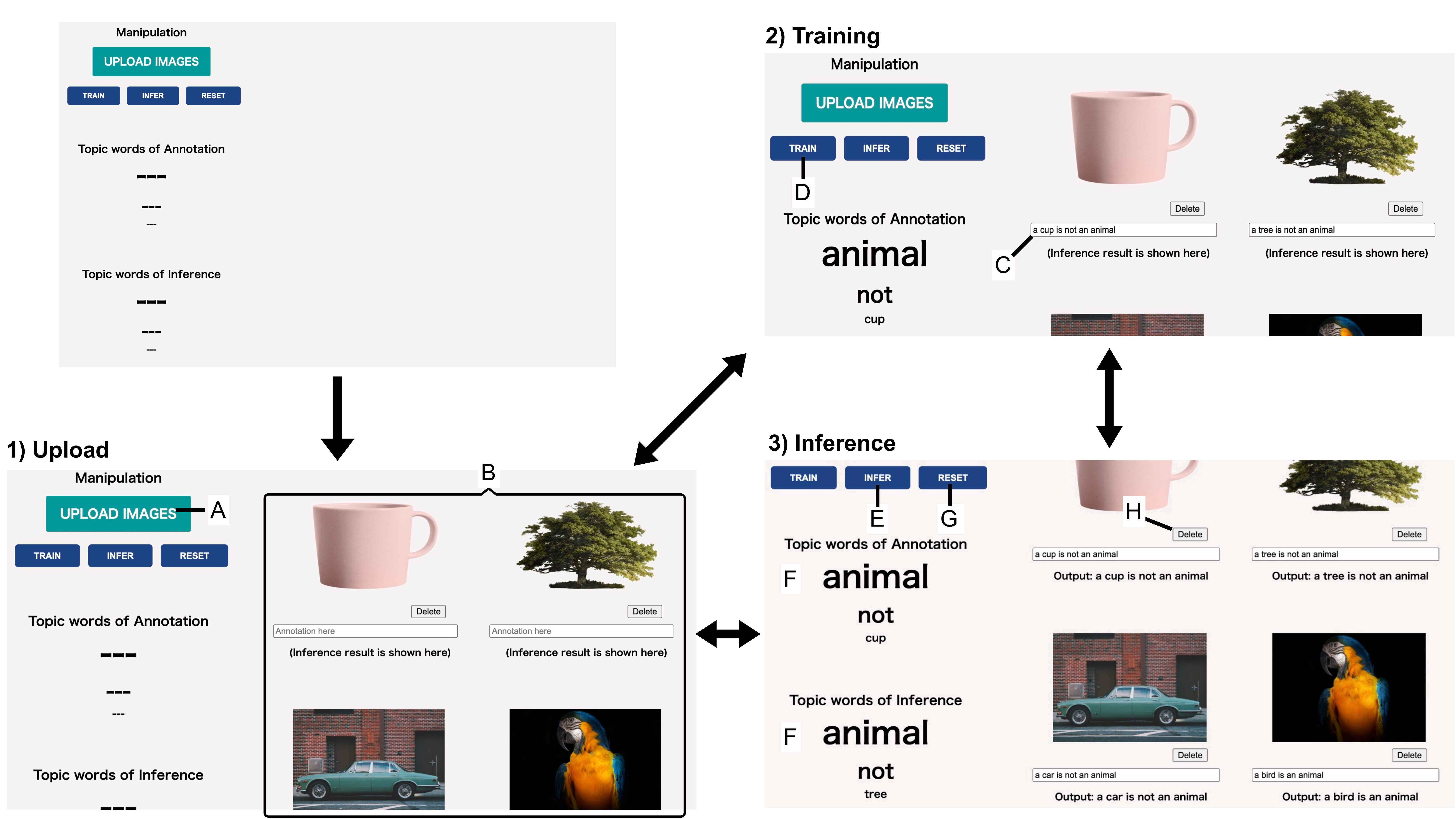}
  \caption{GUI overview of our interactive image-to-text translation system. 1) Users upload images via the upload button (A) and they are displayed in the right panel (B). 2) Users enter sentences into the text boxes (C) below selected images and click the training button (D) to update the model. 3) The inference button (E) shows inference results on all images. Topic words panel (F) shows the frequently appearing words in the annotation or inference results. Users can delete all the uploaded images with the reset button (G), and individual images with the delete button (H).}
  \label{fig:gui_translation}
\end{figure*}

To make a comparison between categorical output (i.e., image classes) and text output (i.e., image descriptions) in IML systems dealing with images as input, we implement two prototypes of interactive image recognition systems.
We introduce the user interface on image-to-text translation and technical details on it and then describe the other based on classification.

\subsection{Interactive Image-to-text Translation}

Our interactive image-to-text translation system is designed to help users train their image recognition models by providing text labels.
The text labels are completely up to users' demands, from complex sentences to simple words, and this design allows users to flexibly formulate recognition tasks with natural language.
Although recent studies indicate that the same goal can be achieved by providing appropriate instructions at the input prompts~\cite{zhou2022learning, zhou2022conditional}, this is not always an easy task for non-expert users.
Our design that fine-tunes the model based on the user-defined input-output pairs has the advantage of making the interaction more intuitive for users.

Figure~\ref{fig:gui_translation} shows the proposed interaction flow.
1) Users first upload candidate images for annotation via the upload button (A), and the images are displayed on the right panel (B).
2) Users then choose some images as training data and enter the desired text labels in the box (C) below each image.
When users click the training button (D), all image-text pairs are used as training data to update the image-to-text translation model.
3) After model training, users can also click the inference button (E) to test the model output on all images, including the unlabeled ones.
This operation is separated from training to avoid extra time for inference.
By displaying the topic words in the annotation and inference results (F), we also expect users to be aware of the bias in the annotations.
Users repeat the training (2) and inference (3) processes to update the model iteratively until it achieves the desired performance.
They can also delete all images by clicking the reset button (G) or individual images by clicking the delete button (H) below the image at any time, as long as the system is not performing training or inference.

\subsection{Implementation Details}
\begin{figure*}[t]
  \centering
  \includegraphics[width=\linewidth]{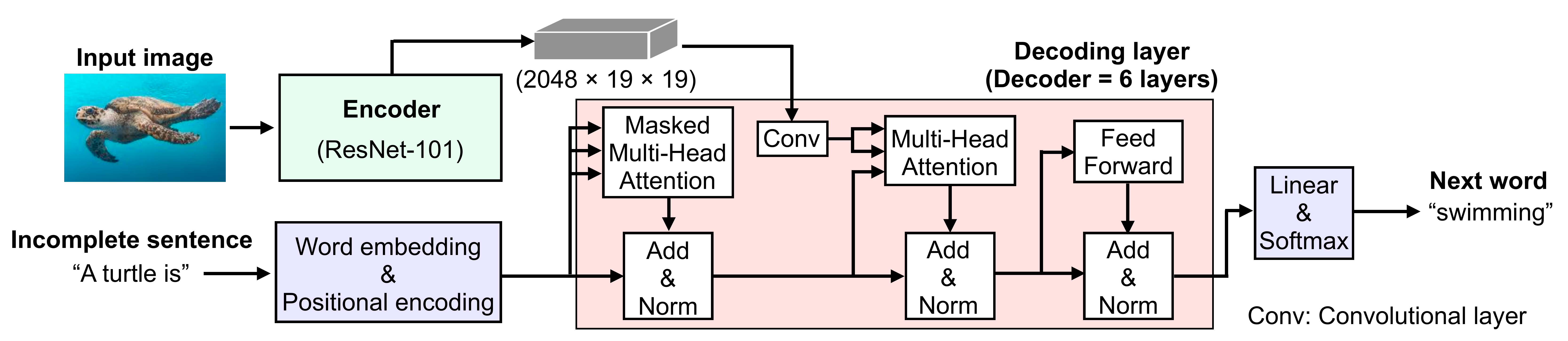}
  \caption{The architecture of the image-to-text translation model consists of CNN-based image encoder and transformer-based text decoder modules. The encoder module takes the input image and encodes the content to the feature tensor. The decoder module works recursively to decode a sentence from the feature one word at a time.}
  \label{fig:architecture_translation}
\end{figure*}

Figure~\ref{fig:architecture_translation} shows the architecture of the image-to-text translation model behind our GUI.
The overall architecture follows previous work on image captioning~\cite{desai2021virtex} and consists of CNN-based image encoder and transformer-based text decoder modules.
The encoder module uses the ResNet-101~\cite{he2016deep} architecture while eliminating the last average pooling layer and the fully connected layer to keep the spatial information.
It takes $299 \times 299$ images as input and extracts $256 \times 19 \times 19$ feature tensor from the input image.
The decoder module uses the same architecture as the decoder part of the transformer~\cite{vaswani2017attention} and consists of six identical decoding layers.
The first decoding layer takes the image feature and the output from the word embedding module with positional encoding as input and generates a feature for the next decoding layer.
Following decoding layers sequentially process the output feature and the image feature, and the last feature output is fed into the linear layer to output the next word.
The output word is then recursively fed into the word embedding module to complete the sentence.

To facilitate frequent training iteration in the interactive scenario, it is important to keep the time required to train the model low.
If we train the whole network including both encoder and decoder modules, it can take about 20 to 30 seconds to fine-tune the model even with a few dozen samples and decent GPUs.
Some prior work on image captioning use pre-trained encoder modules and trains only the decoder module; however, this is still not sufficient to achieve real-time model updates.
In this work, we employ a fine-tuning strategy updating only the last layer of the Transformer decoder.
Specifically, when the user triggers model updates via the training button, only the last layer is updated via backpropagation while the other parameters are fixed.

In our experiment, we implemented the system as a Web app with a JavaScript GUI and Python backend.
Data handling is mostly done on the backend, and GUI communicates with the backend via HTTP requests.
The backend model is pre-trained on large-scale datasets and fine-tuned using user-defined training data.
The encoder is pre-trained on the ImageNet dataset~\cite{deng2009imagenet}, and then the decoder is pre-trained on the MS COCO dataset~\cite{lin2014microsoft} using cross-entropy loss.
To facilitate fast updates, only the last decoding layer is fine-tuned during the interactive training phase.
Each time the user presses the training button, the model is fine-tuned with the cross-entropy loss for 20 epochs always from scratch.
We use the AdamW algorithm~\cite{loshchilov2017decoupled} with batch size $8$, and the learning rate was set at $1e-4$.
When training the last decoding layer, we first calculate and save the output from the fifth decoding layer to further reduce the feed-forward cost.

\subsection{Interactive Image Classification Baseline}

\begin{figure}[t]
    \begin{subfigure}{.49\textwidth}
      \centering
      \captionsetup{width=.95\linewidth}
      \includegraphics[width=\linewidth]{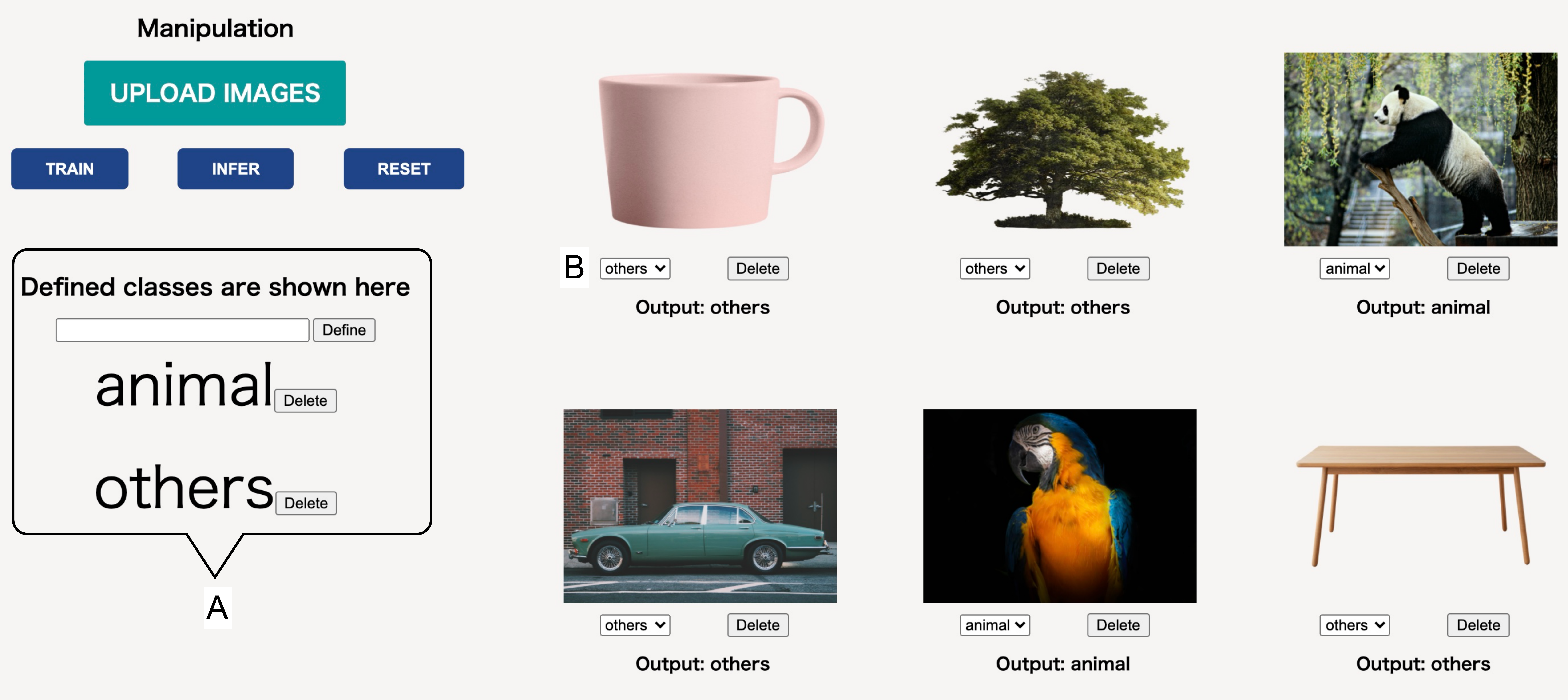}
      \caption{GUI of the classification system. Users first define target categories (A), and select the category from the drop-down list for each image (B).}
      \label{fig:gui_classification}
    \end{subfigure}
    \begin{subfigure}{.49\textwidth}
      \centering
      \captionsetup{width=.95\linewidth}
      \includegraphics[width=\linewidth]{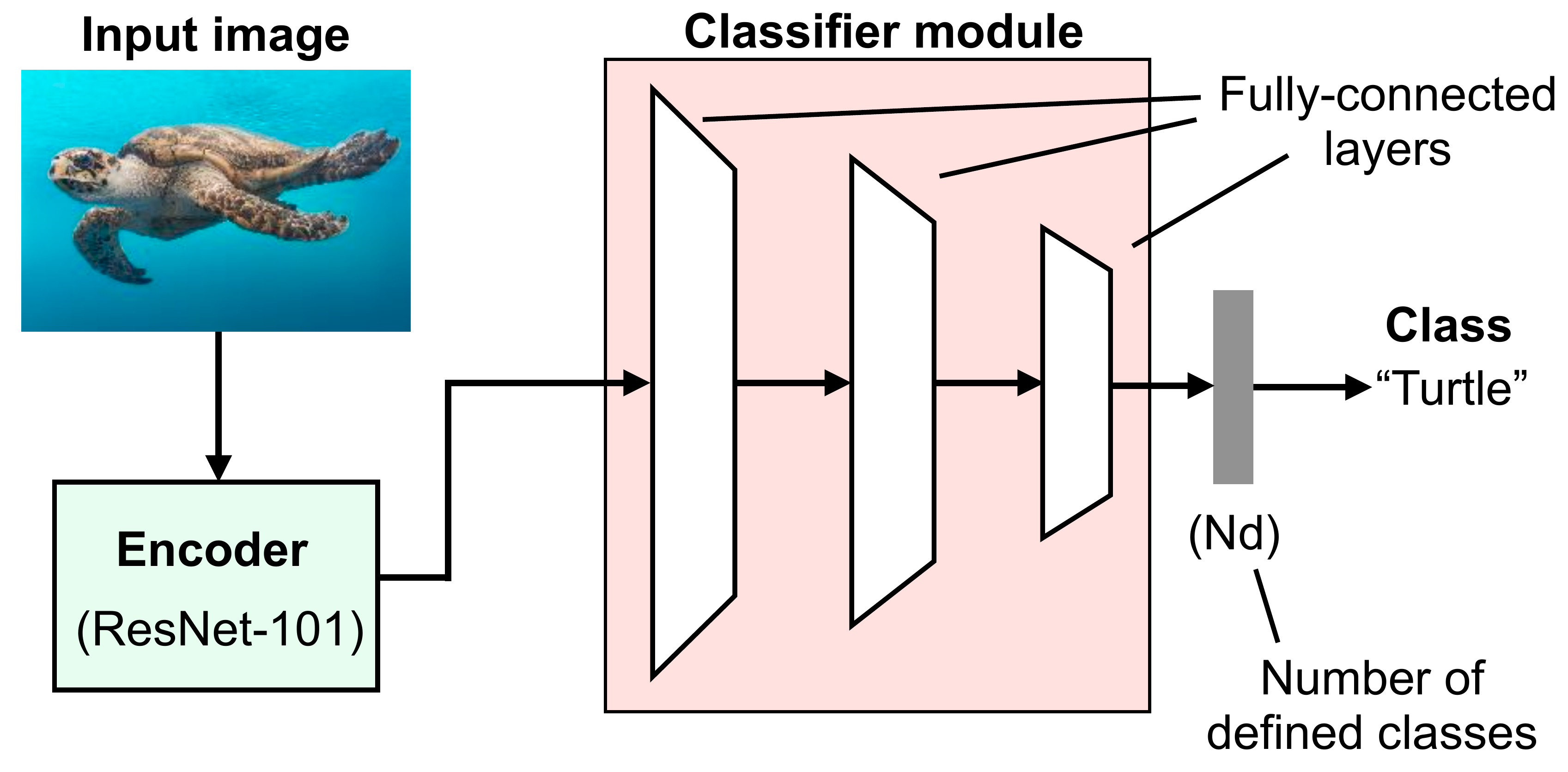}
      \caption{The architecture of the image classification backend consists of the encoder module and classifier modules.}
      \label{fig:architecture_classification}
    \end{subfigure}
  \caption{Overview of the image classification-based system.}
  \label{fig:classification}
\end{figure}

We also implemented an image classification-based prototype whose GUI is designed to be as equivalent as possible to the image-to-text translation system for a fair comparison.
As shown in Fig.~\ref{fig:gui_classification}, users first need to define target categories (A) and select one of them from the dropdown list for each uploaded image (B).
The remaining usage flow stays nearly the same as the image-to-text translation system -- users can freely upload and delete images, trigger model training using the user-annotated pairs, and generate inference results for all uploaded images.

The backend is made up of encoder and classifier modules (Fig.~\ref{fig:architecture_classification}).
The encoder module is based on the ResNet-101 architecture pre-trained on the ImageNet dataset and uses the final $1000$-d vector as the intermediate feature.
The classifier module consists of three fully connected layers and is randomly initialized for each user session.
During interactive usage, the classification module is fine-tuned with user-defined categories.

\section{User Study}

The goal of our user study is to compare two prototype IML systems by letting non-expert participants freely solve image recognition tasks.
Ideally, this study should be done by allowing participants to freely conceptualize their recognition tasks and observe how they formulate them using the IML systems~\cite{nakao2020use}.
However, such a study has the disadvantage of making controlled quantitative comparisons among participants and systems difficult.
Therefore, we opted to simulate this situation by providing participants with rough descriptions of several pre-defined recognition tasks.
We analyze how the participants formulate the given task in each system.

\subsection{Image Recognition Tasks}

\begin{table*}[t]
   \centering
   \caption{Overview of the recognition tasks used in the user study. The tasks can be divided into mainly two categories, \emph{Detection} and \emph{Non-classification}, and each task was presented to the participants together with different source unlabeled image datasets.}
  \label{tab:tasks}
  \begin{tabular}{ccll}
    \hline
    ID & Task   & Goal  & Dataset\\
    \hline
    1  & Detection  & Detecting terrestrial animals & Photo Art 50~\cite{wu2014learning} \\
    2  & Detection  & Detecting people performing exercises & Stanford 40 Action~\cite{yao2011human} \\
    3  & Detection  & Detecting hand-held tools & Caltech 101~\cite{feifei2004learning} \\
    4  & Detection  & Detecting female tops & DeepFashion~\cite{liu2016deepfashion} \\
    5  & Detection  & Detecting meat dishes & Food 101~\cite{bossard2014food}\\
    6  & Non-classification & Estimating the crowd congestion level & CrowdHuman~\cite{shao2018crowdhuman} \\
    7  & Non-classification & Estimating the age of the person & UTKFace~\cite{zhang2017age} \\
    8  & Non-classification & Estimating the size of the area occupied by plants/trees & Stanford Background~\cite{gould2009decomposing} \\
    9  & Non-classification & Estimating the face orientation & AFLW2000-3D~\cite{zhu2016face} \\
    10 & Non-classification & Estimating the ship/boat location & MSCOCO~\cite{lin2014microsoft} \\
    \hline
  \end{tabular}
\end{table*}

As discussed earlier, there are two potential challenges for classification-based design: difficulties for non-experts in setting up abstract categories and the existence of tasks that cannot be formulated as classification.
We prepared ten dummy tasks listed in Table~\ref{tab:tasks} corresponding to these two challenging cases.
The first five tasks correspond to the cases where users need to introduce abstract categories, i.e., categories encompassing multiple specific categories found in the images.
In task 1, for example, the expected text annotation should include category information as ``\textit{this is a terrestrial animal}'' and ``\textit{this is not a terrestrial animal}''.
Annotations that describe specific animal names rather than target categories are considered failure cases.
The last five \textit{non-classification} tasks are the cases that require continuous labels such as regression or segmentation.
The possibility of representing continuous quantities numerically in a language output has been discussed in recent studies~\cite{spithourakis2018numeracy, thawani2021numeracy}.
While it is assumed for \textit{non-classification} tasks that numerical expressions are included in the ideal output text, the participants were allowed to linguistically express the rough quantity. 
For example, in task 8, the participants could also describe the answer as ``\textit{large parts are occupied}'' or ``\textit{almost no plants or trees}'' instead of numerically expressing the percentage of the area.
In both cases, the dummy task descriptions given to the participants were only a use-case scenario and brief task requirements and did not provide specific instructions or concrete examples about the annotation.

Each task was presented to the participants together with different unlabeled source image datasets as shown in Table~\ref{tab:tasks}.
We used existing classification datasets for the former five tasks and randomly selected 40 categories to match the smallest number of categories among all datasets.
Here we assume the scenario where non-expert users use image search engines or equivalent to collect images for annotation.
Images were stored in subfolders with their original category names to simulate the image search and collection process for the development of ML systems.
For the latter five tasks, we randomly selected 200 images from each dataset.
In this case, all images were stored and presented in a single folder.
In both cases, all image files were renamed as sequentially numbered files in random order.

\subsection{Procedure}

We recruited 20 (11 female) participants ranging from 21 to 48 (M $=35.95$, SD $=8.73$) years old through a staffing agency.
None of the participants had any knowledge about how ML works and only two of them answered that they had experience using ML applications.
At the beginning of the study, we first provided a brief explanation of the concept of interactive machine learning and the usage of both systems.
Afterward, the participants spent 10 minutes getting familiarized with both systems by solving a simple recognition task (animal classification).
In the main study, the participants solved four randomly assigned tasks, two with the image classification system and the rest with the image-to-text translation system.
Of the four tasks, two were detection tasks and two were non-classification tasks.
Note that the task-system combinations are counterbalanced, and the order of tasks and systems was randomized for each participant. 
We set the maximum time for each task as 20 minutes while allowing participants to finish the trial when they felt that the annotation was sufficiently done.
The application server was running on a GPU-equipped workstation, and the participants accessed the Web interface using the same Web browser from the same model laptop PCs.
We also instructed them to use the same online machine translation website as a reference for English translation.
We recorded all user interaction logs during the study.

After all trials, the participants also answered the subjective questionnaire.
They first answered the following six questions on a 7-point semantic differential scale for each system:
\begin{description}
    \item[PQ1] The system was, in general, easy to use 
    \item[PQ2] The system allowed intuitive operation
    \item[PQ3] You could create training data efficiently
    \item[PQ4] You could include the necessary information in your training data
    \item[PQ5] The way of creating training data (category label or text) was effective for training 
    \item[PQ6] The system could be useful in everyday situations
\end{description}
They also answered the NASA-TLX test~\cite{hart1988development} to rate the workload.
We adopted the weighted score while excluding the \emph{physical demand} score.
Lastly, to understand their in-depth thoughts, we asked them to complete an open questionnaire on the aspects to which they paid attention during the training processes.

Since it is impossible to automatically judge the appropriateness of the user-defined tasks, we opt for third-party evaluation to evaluate annotation results and model performances.
We asked eight graduate students engaged in computer vision research and asked them to subjectively judge how the participants addressed the tasks.
The evaluation was done by answering the following questions on a 5-point semantic differential scale for the result of each task trial:
\begin{description}
    \item[EQ1] Category labels or text annotations satisfy the requirements of the task
    \item[EQ2] Training images are appropriately selected to train the model well 
    \item[EQ3] The trained model works as intended by the participant 
\end{description}
EQ1 and EQ2 are about training data, and we asked the questions together with image-annotation pairs created by the participant.
EQ3 is about inference results and introduced another test data for each task and presented pairs of test images and their inference results to evaluators.
We assigned evaluators so that each task trial has three evaluators, and use the median scores for the following analyses.

\subsection{Results}

\subsubsection*{Annotation Analysis}

\begin{figure}[t]
  \begin{minipage}[b]{0.48\linewidth}
    \centering
    \includegraphics[keepaspectratio, scale=0.2]{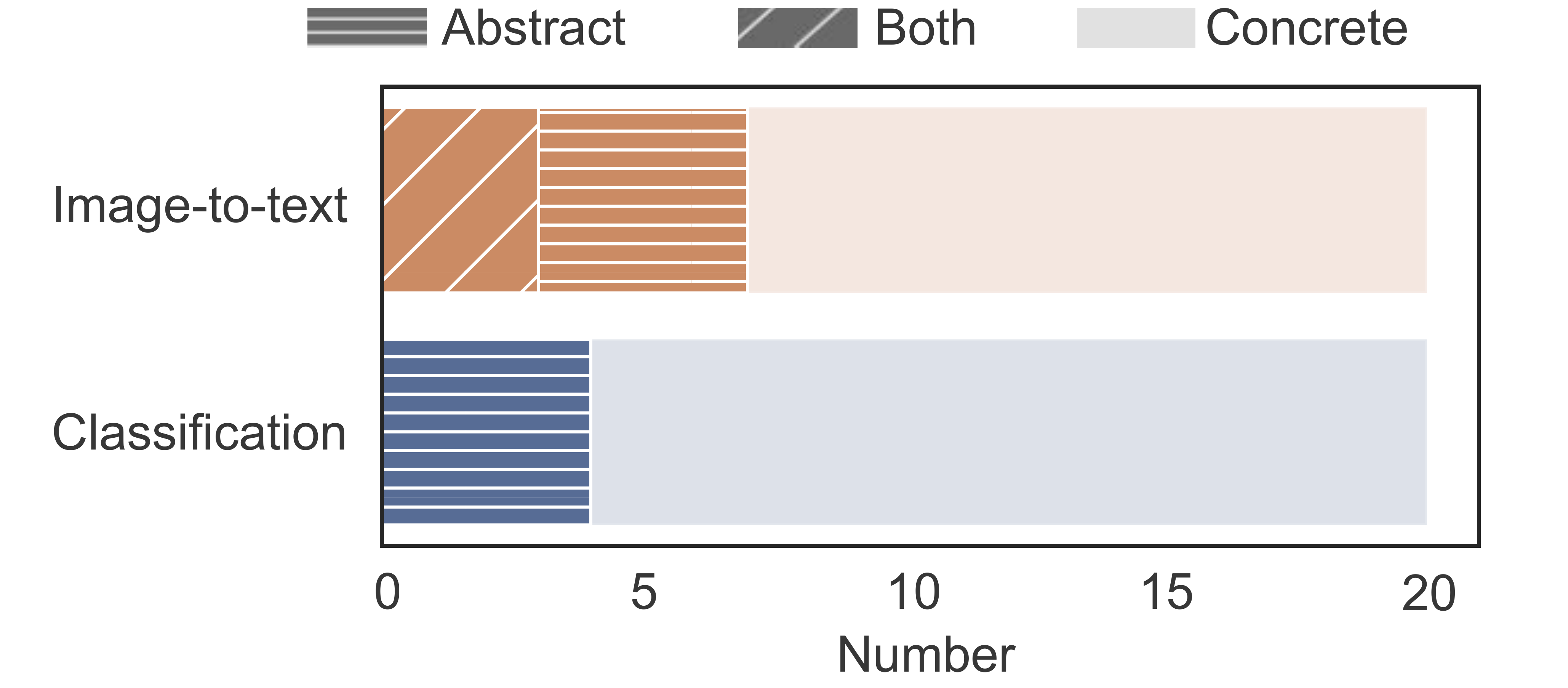}
      \subcaption{The proportion of task trials in the detection category introduced abstract categories in the annotations}
      \label{fig:stats_det}
  \end{minipage}
  \begin{minipage}[b]{0.48\linewidth}
    \centering
    \includegraphics[keepaspectratio, scale=0.5]{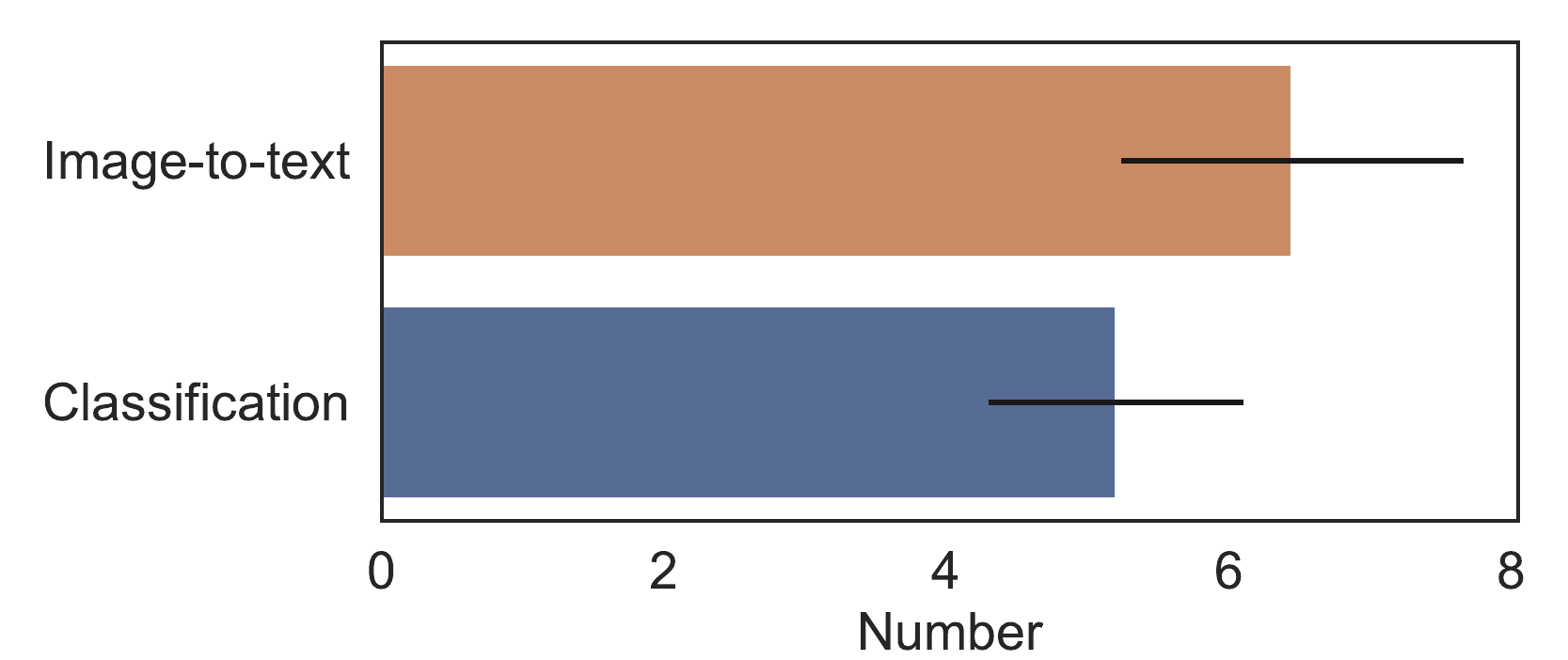}
      \subcaption{The average label granularity for the tasks in the non-classification category. Error bars indicate standard errors.}
      \label{fig:stats_reg}
  \end{minipage}
  \label{fig:stats}
    \caption{Annotation statistics for the tasks in the (a) detection and (b) non-classification category.}
\end{figure}

Figure~\ref{fig:stats_det} shows the number of task trials in the detection category that introduced abstract category labels.
As discussed earlier, abstract category labels are required to properly formulate these tasks.
The shaded part corresponds to the case in which concrete category names are included together with the abstract category.
With the image classification system, the participants did not introduce abstract categories and assigned concrete object names (``\textit{Kangaroo}'', ``\textit{Anchor}'', ``\textit{PC}''...) as category labels in most cases (16 of 20).
Although there are still many failure cases, we observed more cases (7 out of 20) mentioning the concept of abstract categories in the annotated text with the image-to-text translation system.
This includes the cases where annotations include both concrete categories and abstract concepts.
For example, one participant entered texts like ``\textit{Brushing your teeth is not a sport}'', ``\textit{Playing the violin is not a sport}'', and ``\textit{Running is a sport}'' for the task 2.
In this case, the names of the specific action (i.e., brushing, playing, running) are concrete categories and the noun phrases that encompass them (i.e., sport, not a sport) are abstract categories.

Figure~\ref{fig:stats_reg} shows the average label granularity introduced by the participants for the quantity expression, i.e., how finely/detailed the user divided and labeled the target quantities in each trial.
We counted how many different ways the participants phrased a particular quantity.
For example, if the crowd was described in three ways: ``\textit{crowded}'', ``\textit{somewhat crowded}'', and ``\textit{uncrowded}'' in task 6, we counted a granularity as three.
The image-to-text translation system resulted in higher granularity (M$=6.45$) than the image classification system (M$=5.20$), and the participants described the details of the quantities more precisely with the image-to-text translation system.
For example, a participant defined the location category for task 10 with a single word such as ``\textit{right}'', ``\textit{left}'', and ``\textit{center}'' with the image classification system; another participant with the image-to-text translation system described directions with more detailed expressions like ``\textit{slightly right of center}'', ``\textit{lower left}'' or ``\textit{left-upper center}''.

\begin{figure}[t]
  \begin{minipage}[b]{0.48\linewidth}
    \centering
    \includegraphics[keepaspectratio, scale=0.32]{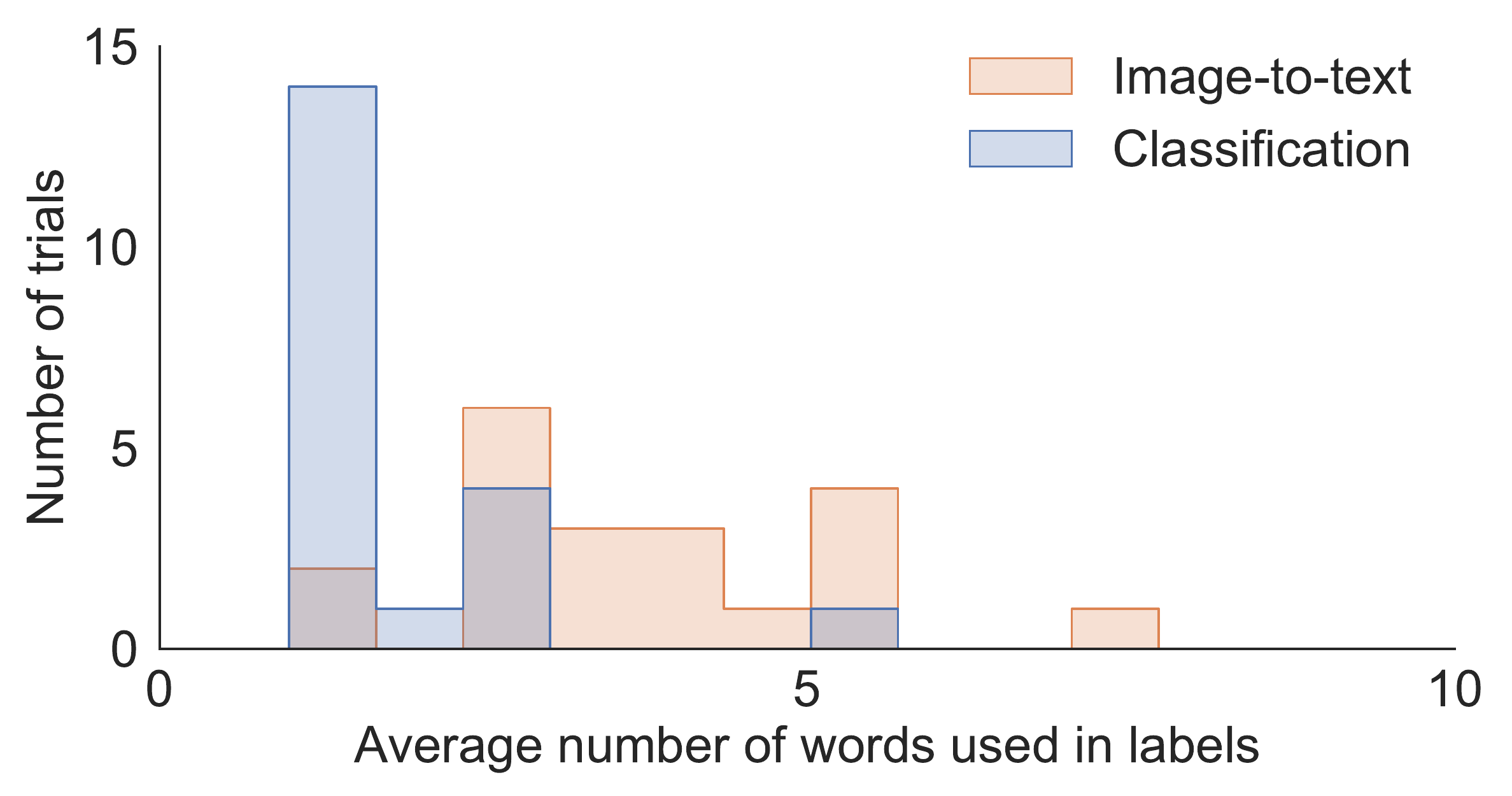}
      \subcaption{Detection tasks.}
      \label{fig:words_det}
  \end{minipage}
  \begin{minipage}[b]{0.48\linewidth}
    \centering
    \includegraphics[keepaspectratio, scale=0.32]{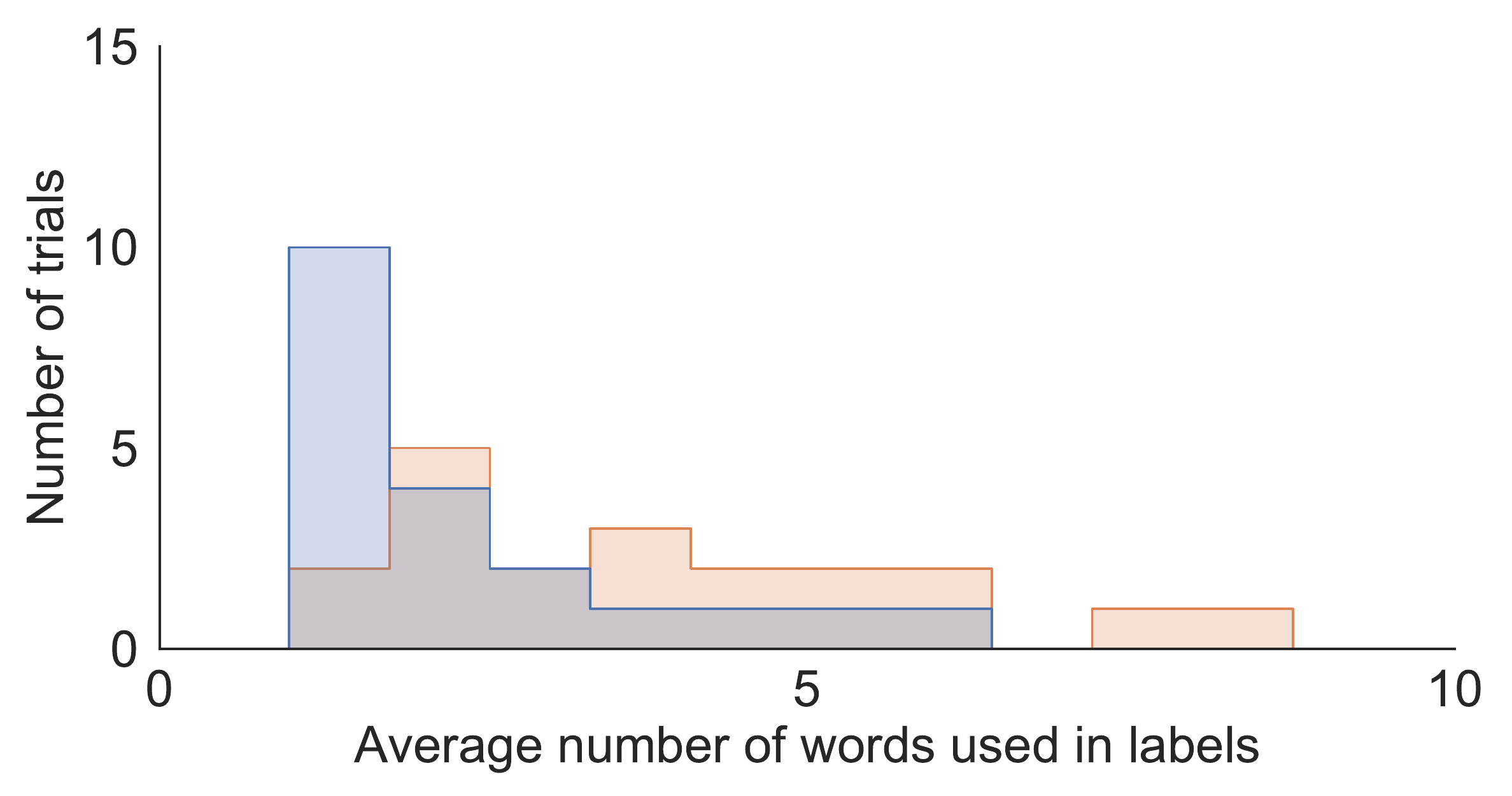}
      \subcaption{Non-classification tasks.}
      \label{fig:words_reg}
  \end{minipage}
  \caption{Distributions of the average number of words used in annotations for each task trial.}
  \label{fig:words}
\end{figure}

Furthermore, Figure~\ref{fig:words} shows the distribution of the average number of words included in the annotation labels.
While it is natural that the annotations given with the image-to-text translation system included significantly more words (Mann-Whitney U test, $p < 0.01$), we can observe that the participants provided long sentence labels even for the classification system.
This indicates the tendency of participants to include unnecessarily detailed information in the labels.
In the user questionnaire, one participant commented that s/he tried to provide ``\textit{detailed description of the images}'' while using the image classification system.
Another participant also commented that s/he intentionally used ``\textit{words and proper nouns expected to be important for the target task.}''

\subsubsection*{Subjective Evaluation}

\begin{figure*}[t]
    \begin{subfigure}{\textwidth}
      \centering
      \captionsetup{width=.95\linewidth}
      \includegraphics[width=\linewidth]{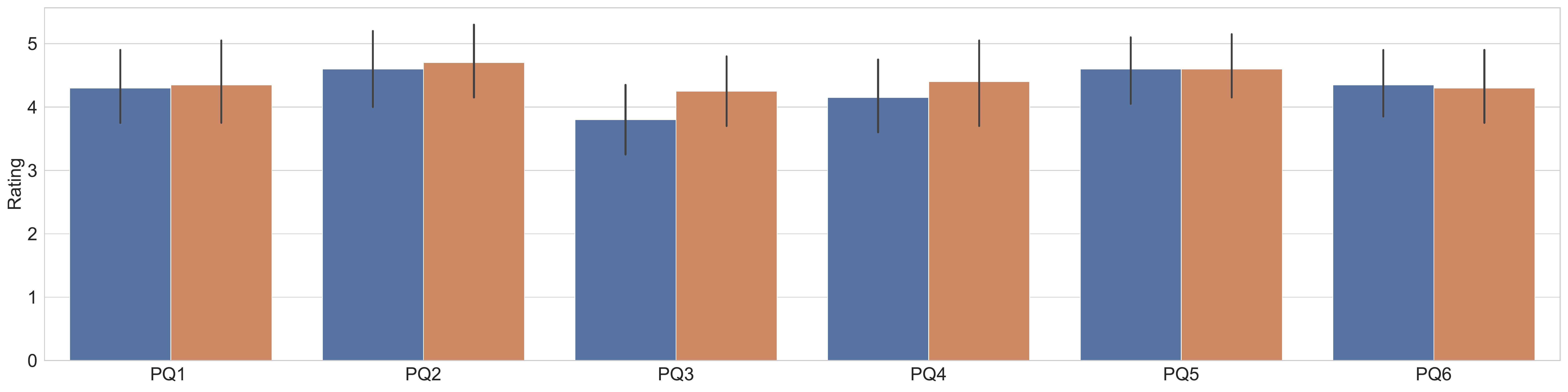}
      \caption{Scores of the usability questionnaire.}
      \label{fig:usability}
    \end{subfigure}
    \begin{subfigure}{\textwidth}
      \centering
      \captionsetup{width=.9\linewidth}
      \includegraphics[width=\linewidth]{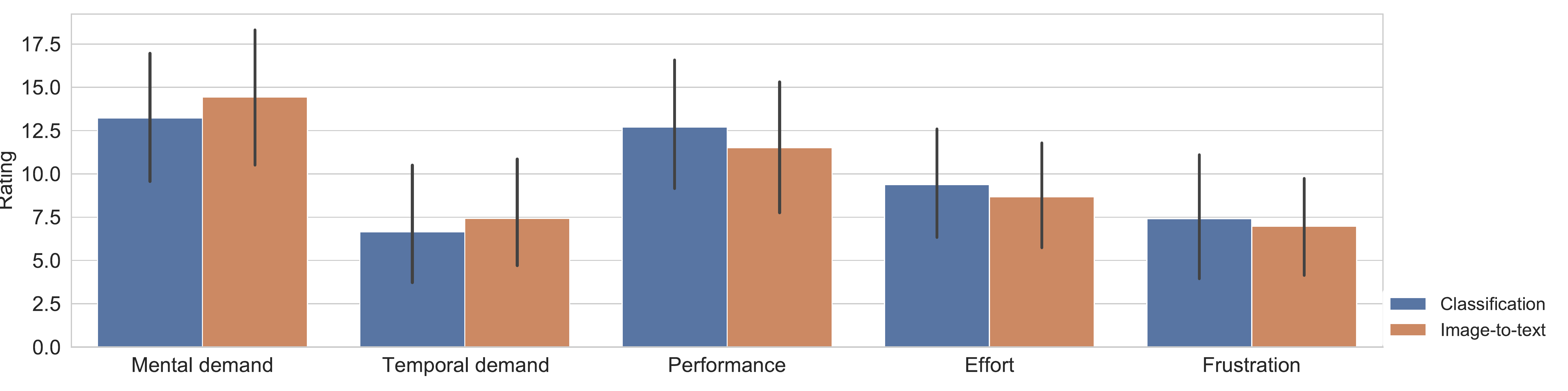}
      \caption{NASA-TLX scores.}
      \label{fig:tlx}
    \end{subfigure}
  \caption{Results of the subjective evaluation. Each block shows the average score and the error bars show their standard deviations.}
  \label{fig:usability_tlx}
\end{figure*}

Figure~\ref{fig:usability} summarizes the usability questionnaire of the two systems.
Similarly, Figure~\ref{fig:tlx} summarizes the NASA-TLX scores for the two systems.
Each block shows the average score for each question and the error bars show their standard deviations.
The left and right bars correspond to the classification and image-to-text translation systems, respectively.
There were no significant differences between the two systems on any of the items.
Despite the fact that the image-to-text translation system requires users' efforts to type annotation texts, there was no significant loss of usability compared to the classification baseline.

Regarding the creation of training data, two participants mentioned that they tried to annotate various images using the image classification system.
One participant stated: ``\textit{Even when choosing multiple images from the same category, I tried to let the model learn that there are various patterns}.''
Another participant commented on the discrepancies between categories (although this strategy prevents learning difficult cases): ``\textit{I tried to emphasize the differences between categories. 
For example, in Task 5 about meat dish detection, I chose brown images for meat dishes and colorful images for dishes that are not meat}.''
Although the choice of the word does not directly influence the model training process, three participants noted that they explicitly included words that would be helpful in solving the task: ``\textit{I included words related to the task when I created category labels}'', ``\textit{I included the most important words}.''

Perspectives on the amount and diversity of training data were also observed for the image-to-text translation system.
A participant mentioned the balance of image selection: ``\textit{I consciously tried to train the model in a balanced manner. For example, if I annotate three images of non-living things in a row, then I provide three images of creatures.}''
Another participant mentioned the style of the images: ``\textit{I selected both illustrations and photos as training data so that the model can correctly classify whichever input is given.}''
The participants also commented on how to write appropriate texts.
Six participants mentioned that they tried to keep the texts as simple and clear as possible.
Examples of typical comments are as follows: ``\textit{I tried to create training data with easy-to-understand words and concise sentences}'', and ``\textit{To minimize the differences in grammar and vocabulary, I tried to use the same words for similar expressions.}''

The common impression of both systems for the participants was the difficulty of achieving high performance.
Five and three participants mentioned such an impression for classification and image-to-text translation, respectively, slightly more for the classification system.
One participant commented on the classification system: ``\textit{The system itself was very easy to understand and get used to. However, there were times that the rationale behind the result was unclear and I was unable to train the model as I wanted.}''
Another participant also commented on image-to-text translation as: ``\textit{I could not train the model as much as I wanted. I wanted to examine what kind of text input would be appropriate.}''
For the image-to-text translation system, there were also some comments about the training procedure.
One participant noticed that the inference results depend on how and which part of the image they described in the annotation text: ``\textit{I noticed that the inference results change depending on which part of the image is described.}''
The fact that users care about the format of annotation texts is a unique aspect of the image-to-text translation system and is not observed in the classification system.

\subsubsection*{Third Party Evaluation}
\begin{figure*}[t]
    \begin{subfigure}{0.33\textwidth}
      \centering
      \includegraphics[width=\linewidth]{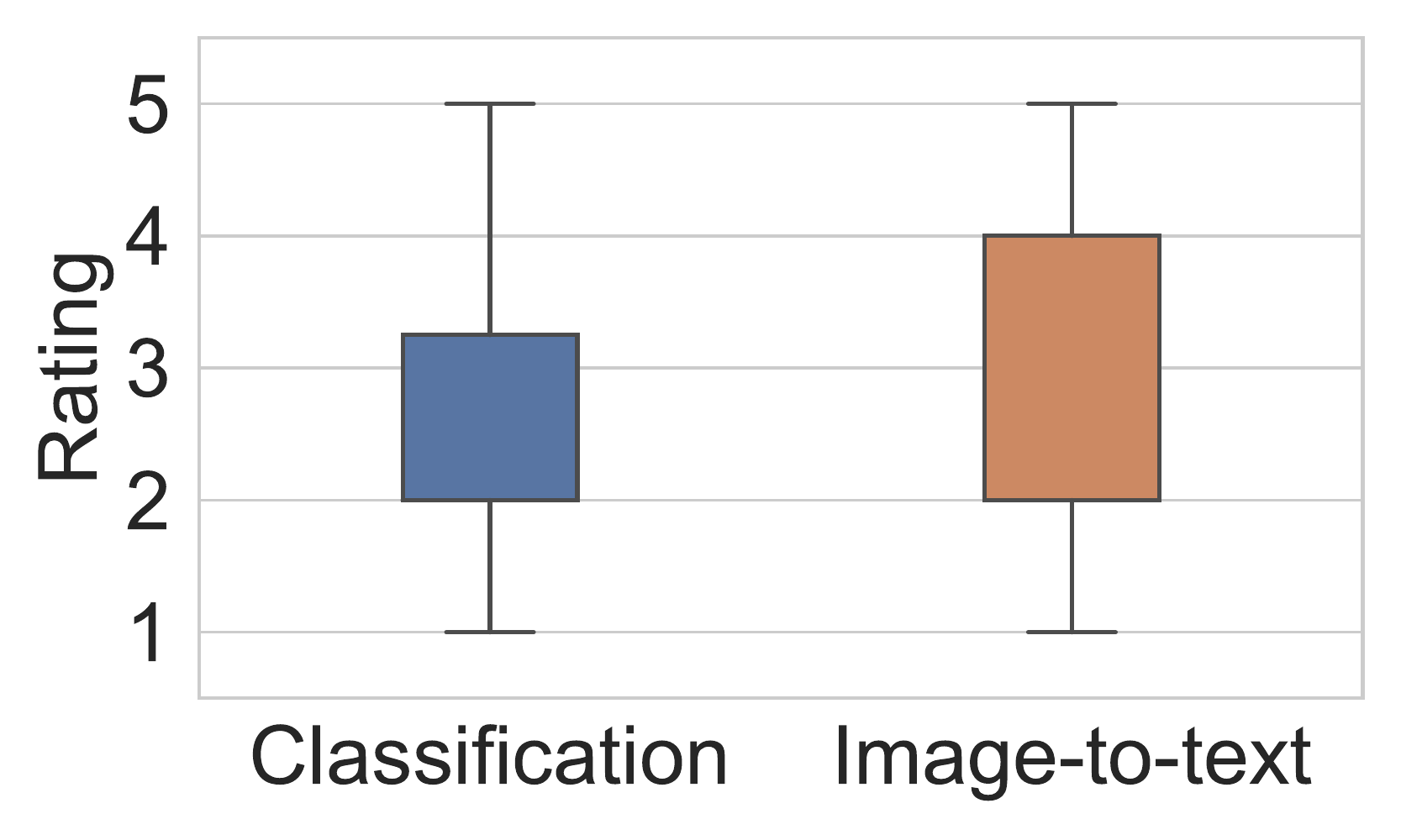}
      \caption{EQ1 (Detection)}
      \label{fig:eq1_det}
    \end{subfigure}
    \begin{subfigure}{0.33\textwidth}
      \centering
      \includegraphics[width=\linewidth]{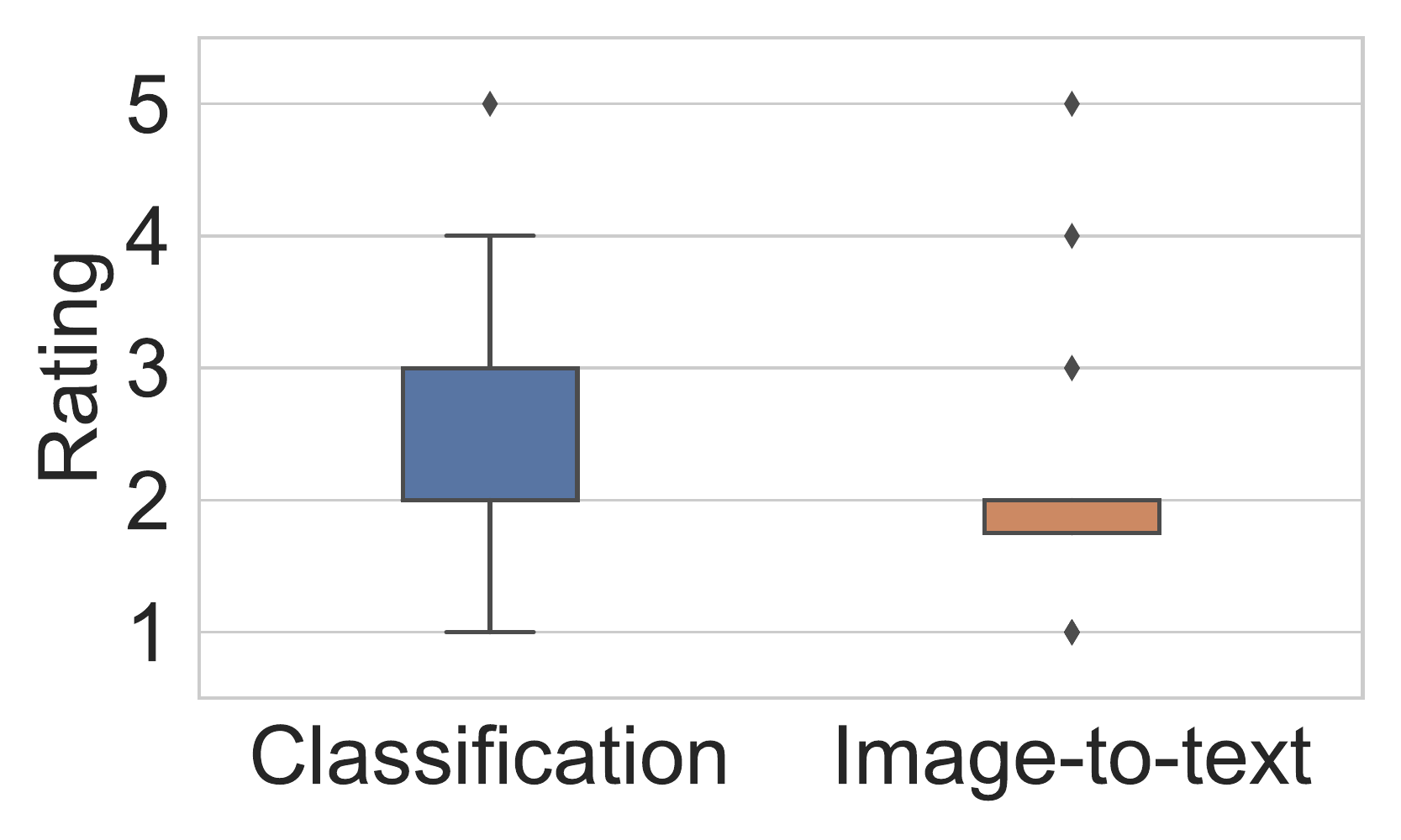}
      \caption{EQ2 (Detection)}
      \label{fig:eq2_det}
    \end{subfigure}
    \begin{subfigure}{0.33\textwidth}
      \centering
      \includegraphics[width=\linewidth]{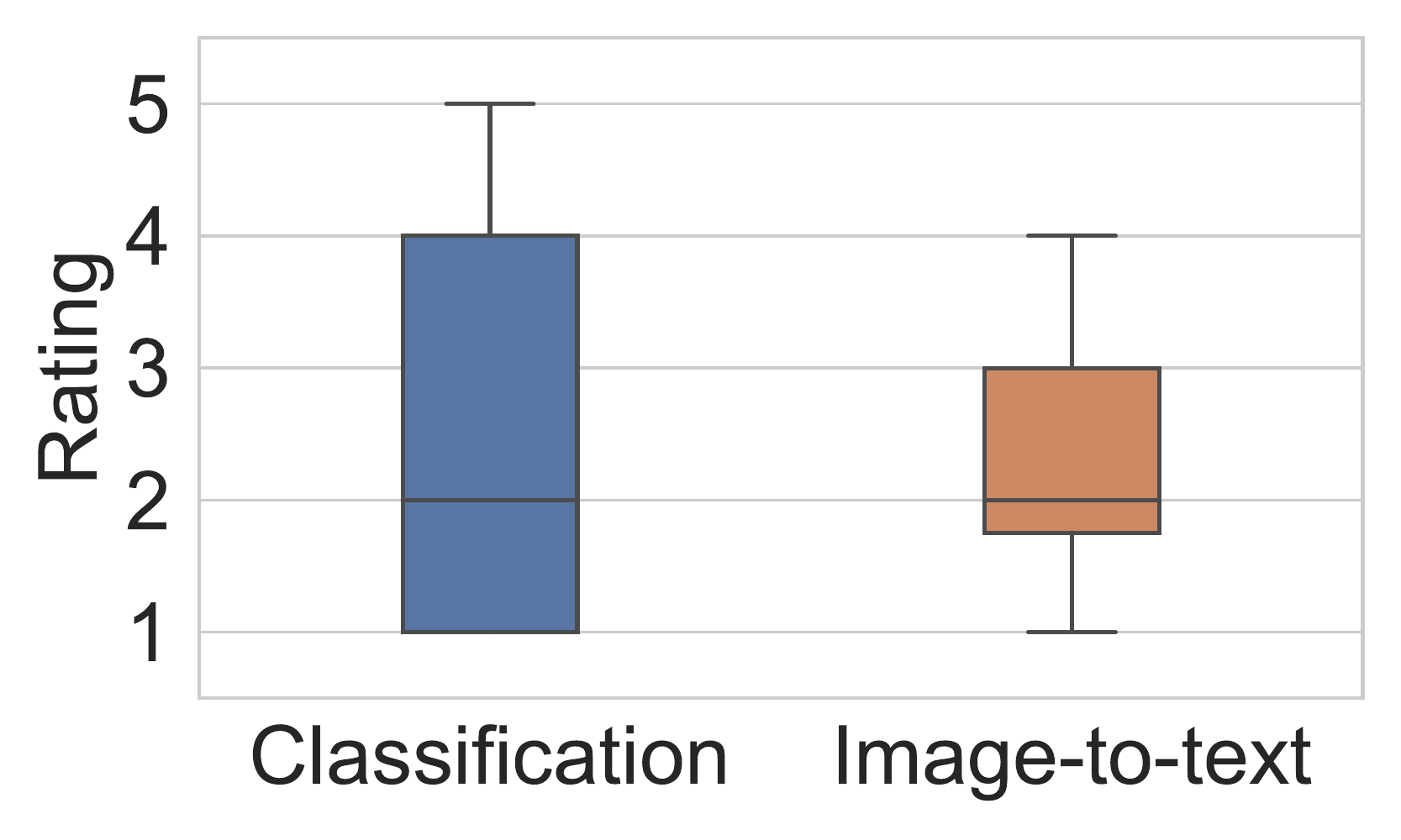}
      \caption{EQ3 (Detection)}
      \label{fig:eq3_det}
    \end{subfigure} \\
    \begin{subfigure}{0.33\textwidth}
      \centering
      \includegraphics[width=\linewidth]{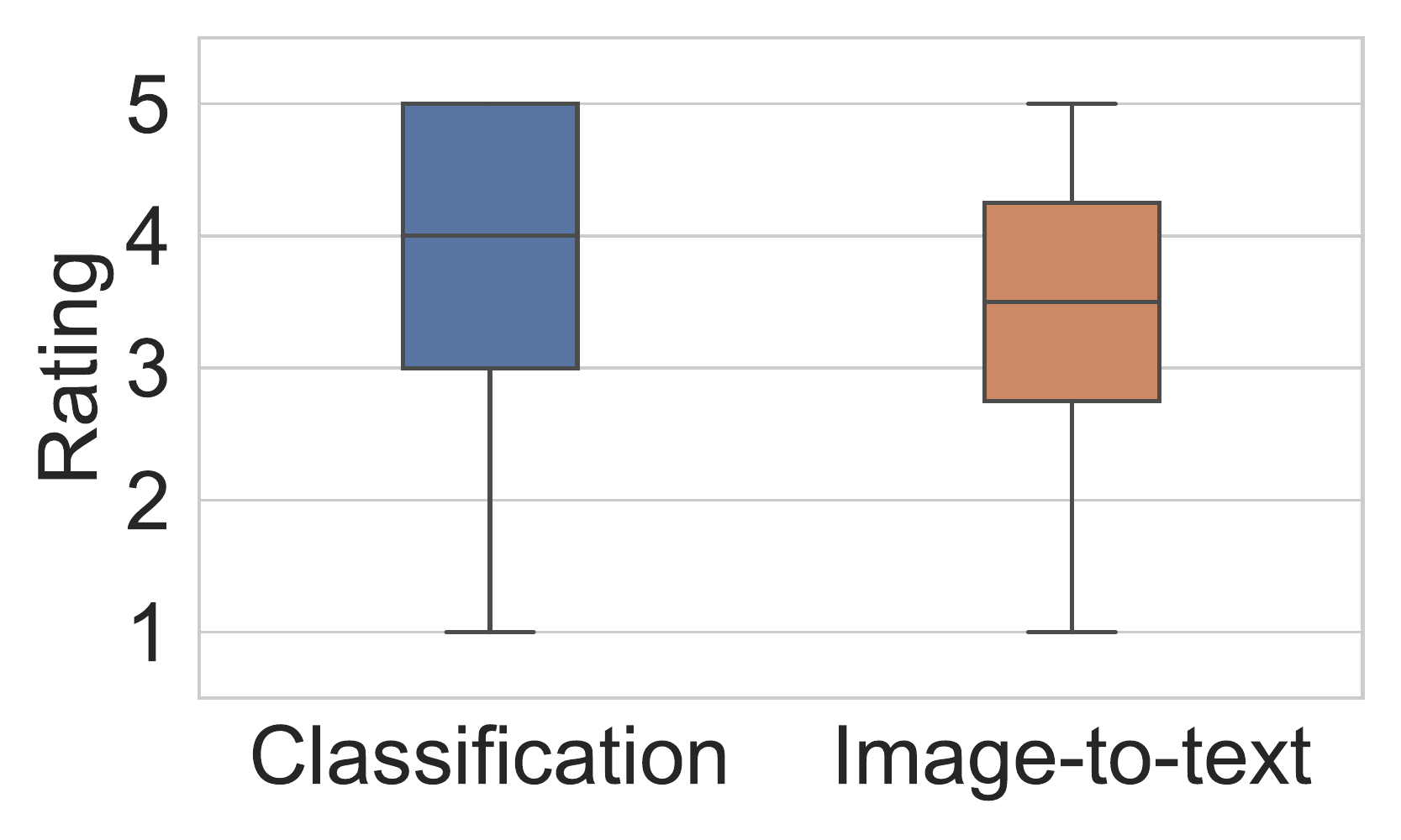}
      \caption{EQ1 (Non-classification)}
      \label{fig:eq1_reg}
    \end{subfigure}
    \begin{subfigure}{0.33\textwidth}
      \centering
      \includegraphics[width=\linewidth]{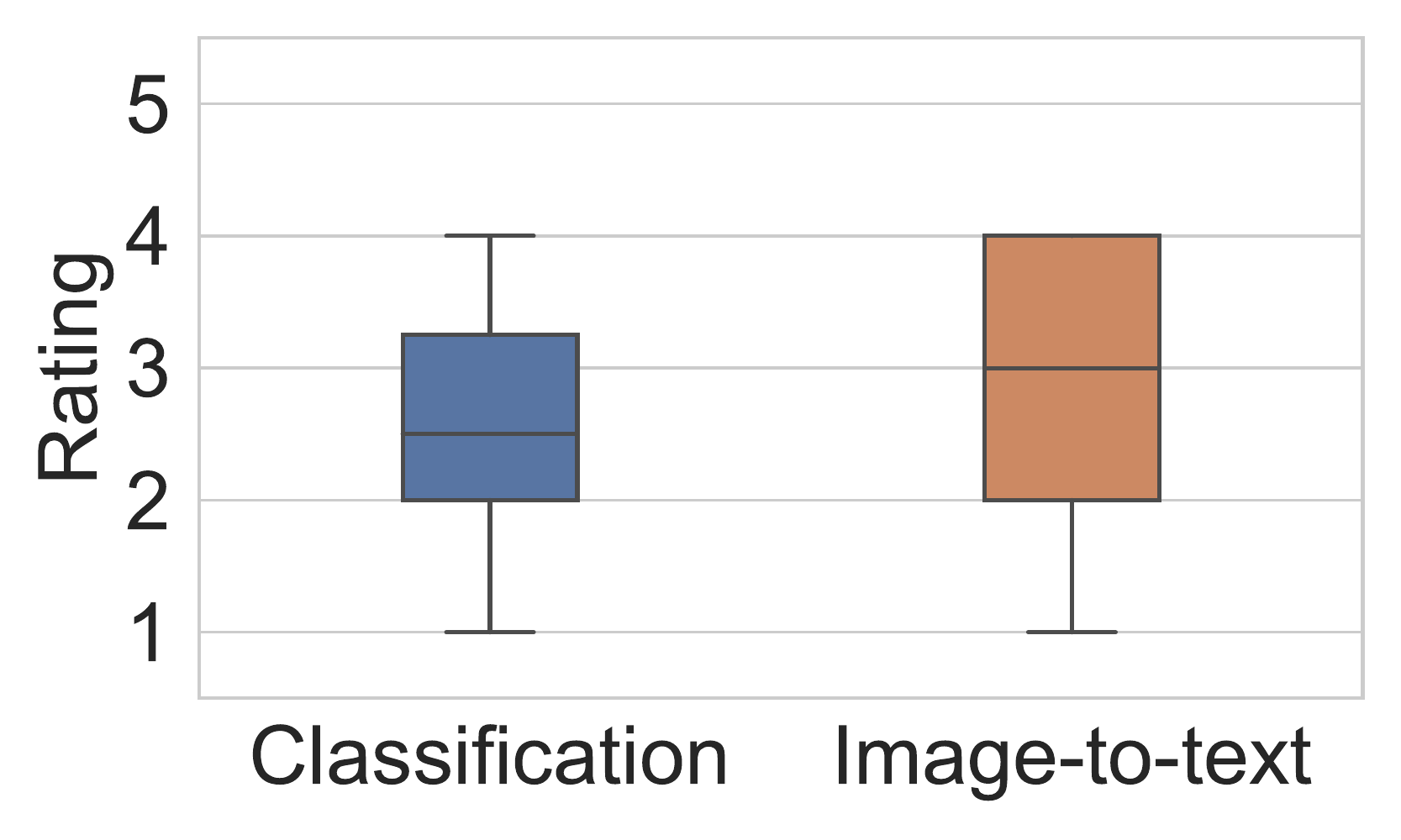}
      \caption{EQ2 (Non-classification)}
      \label{fig:eq2_reg}
    \end{subfigure}
    \begin{subfigure}{0.33\textwidth}
      \centering
      \includegraphics[width=\linewidth]{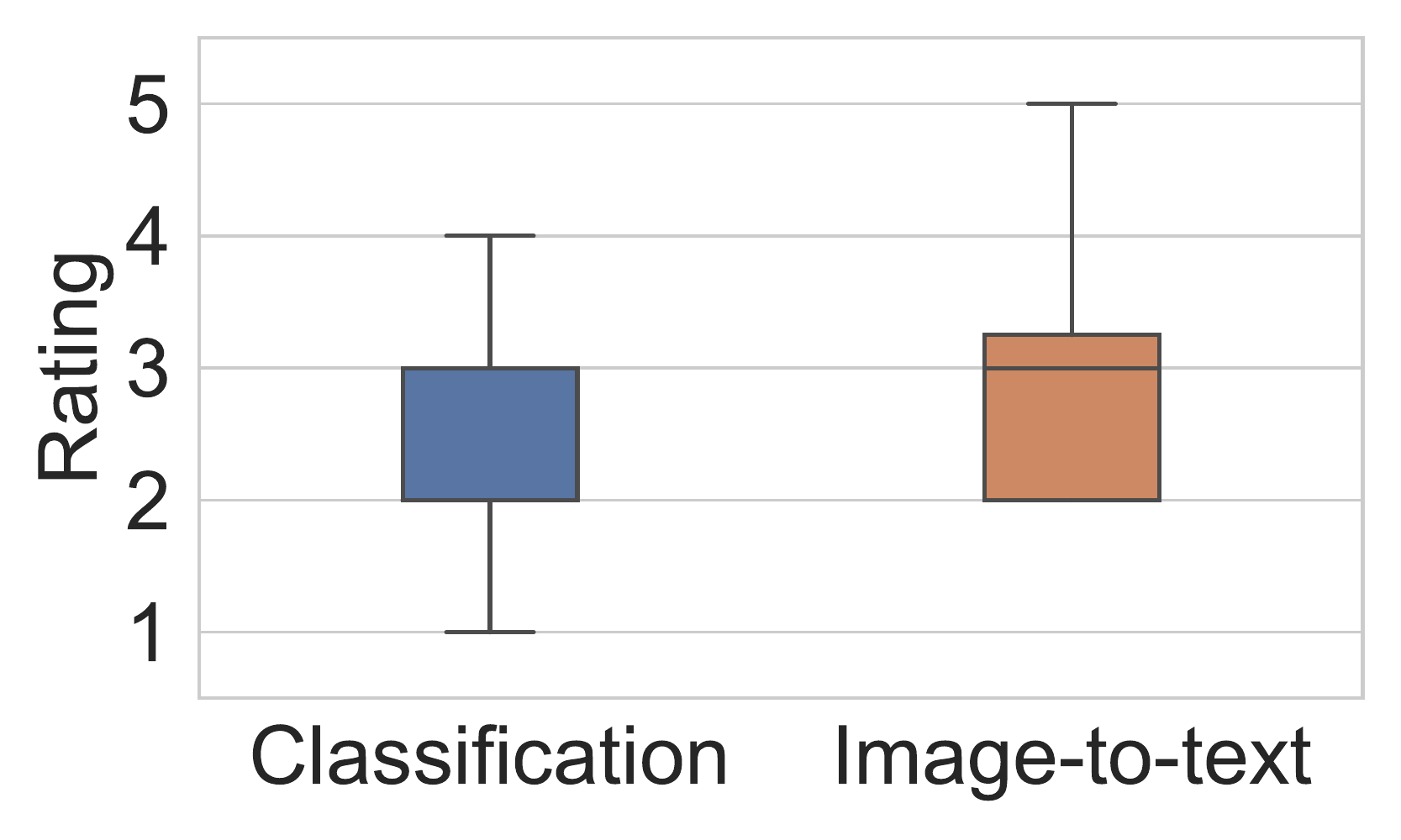}
      \caption{EQ3 (Non-classification)}
      \label{fig:eq3_reg}
    \end{subfigure}
  \caption{Distribution of third-party evaluation scores. The plots correspond to the image classification and image-to-text translation systems from left to right, respectively.}
  \label{fig:evaluation}
  \centering
\end{figure*}

Figure~\ref{fig:evaluation} shows the distribution of third-party evaluation scores.
Each plot shows the distribution of the scores for the two systems.
Regarding EQ1 about the correctness of the user annotations, the image-to-text translation achieved overall better scores than the classification baseline.
EQ2 about training data selection was compared on detection tasks, but slightly more cases were observed on regression tasks with image-to-text translation.
On Non-classification tasks, EQ3 resulted in significantly higher scores with the image-to-text translation system (Mann-Whitney U test, $p = 0.049$).
Despite the invariant quality of annotation and training data creation, the image-to-text translation system generated more reliable outputs for the Non-classification tasks.

\section{Discussions}

In this section, we summarize the key findings from our study and discuss future implications for designing interactive image-to-text translation systems.

\subsection{Key Findings}

In the user study, we investigated the possibility of non-expert users making an image recognition model for various tasks using two systems.
The notable tendency is that correctly defining the task formulation is difficult even with image-to-text translation, though there are possibilities and/or limitations particular to each algorithm.

\paragraph{Challenges in Classification-based Design}
We confirm the difficulty of participants in correctly defining abstract category labels with the classification system.
Only four of 20 cases defined abstract categories and this supports the trend reported in previous work~\cite{nakao2020use}.
While some participants succeeded in defining labels that satisfy the requirements for tasks that require numeric annotation, the granularity is lower than those of the image-to-text translation system.
In general, there was a tendency to simplify the task itself by defining categories that represent quantities with simple words or numbers.
This might be due to the nature of the classification task, which requires mapping a discrete label to each image.

\paragraph{Potential of Text Output in IML}
The image-to-text translation system resulted in seven cases that included abstract descriptions of 20 in total, most notably the cases describing both concrete and abstract categories.
Although we observed that text output cannot fully resolve the difficulty of defining abstract categories, this suggests the potential of natural language description for interactive image recognition.
We also observed that participants described sentences with richer and more fine-grained expressions for annotating continuous quantities in the non-classification category.
This also suggests that users may be able to create training data with text output while maintaining the underlying complexity of the target task.
Furthermore, according to the usability and the NASA-TLX scores, there were no significant differences in the subjective usability of the two systems.
This suggests that, at least from a subjective perspective, text output has the potential to provide user-friendly interfaces.

\paragraph{Difficulties in Handling Semantics with IML}
Our study suggests that the difficulty for non-expert users is not in the output format or the design of the algorithm but in more essential aspects of ML understanding.
The average word lengths of the labels are generally long (Fig.~\ref{fig:words}) even for the classification system, and this fact implies that some participants understand that the \emph{semantics} of the label itself influences the model training process.
Users also tend to think that a larger and more detailed vocabulary trains better recognition models, even if it is not directly related to the task.
This implies that non-expert users have mistakenly assumed that the information contained in the text itself would help make the image recognition model smarter.
This gap between the use of natural language as an interface and its underlying mathematical meaning is one of the fundamental issues that prevent users from technical understanding.

\subsection{Limitations and Future Work}
The inefficiency of annotation is the most critical weakness of the IML systems with text output. 
At least in the proposed system, users have to write down text one by one for each image, and this process is obviously inefficient compared to the classification baseline.
Although many existing interactive classification systems allow users to assign multiple training samples at once to the target class, it is not trivial to achieve such functionality with the output interface of text.
As the use of language becomes more dominant in recent ML technologies, it is an important future work to improve the interaction between users and text-related ML algorithms in general.

Another fundamental limitation of text output is that it is still difficult to actually solve diverse recognition tasks accurately.
For example, even if users provide fine-grained numerical annotations for tasks requiring continuous labels, it is not clear whether the decoder can solve such regression-like tasks by interpolating training samples in the continuous output space.
The decoder relies on pre-training, and it is also difficult to obtain a generic image representation that can be applied to arbitrary tasks.
In the image-to-text translation system, the current backbone CNN is trained on the object-centric ImageNet dataset, and the intermediate feature might not be the best option to solve non-object images such as faces (Task 9) and pixel-wise segmentation (Task 8).
Together with the GUI, it is also important to explore the backend architecture that is capable of solving various recognition tasks.

Although this study was conducted with completely novice users recruited for a paid experiment, it is also important to conduct experiments that take into account the different attributes and motivations of the users.
Domain-expert users, for example, will potentially show different tendencies, and user behavior will also be different if the users themselves are highly motivated to learn and use ML technologies in their daily lives.
As in the case of classification-based systems, long-term, and more real-life user analysis will provide further insight into IML systems with text output.

\section{Conclusion}

In this paper, we report on our prototype study that compares categorical output and text output in IML systems, treating images as the subject matter.
By conducting a user study with non-expert participants, we investigated how they created training data for various image recognition tasks.
The results demonstrated the potential of using natural language output for interactive image recognition, as well as the tendency observed in previous research that employs classification. 
On the other hand, our results also clarified the intrinsic difficulty for non-expert users to accurately understand and imagine ML concepts and formulations.
To bridge the gap between a semantic-driven GUI and a mathematical background, our future work includes developing more efficient and intuitive systems that use natural language as an interface with the user of all attributes.

\section*{Acknowledgments}
This work was supported by JST CREST Grant Number JPMJCR19F2, Japan.

\bibliographystyle{unsrt}  
\bibliography{references}  

\begin{thebibliography}{10}

\bibitem{fails2003interactive}
Jerry~Alan Fails and Dan~R Olsen~Jr.
\newblock Interactive machine learning.
\newblock In {\em Proceedings of the international conference on Intelligent
  User Interfaces}, pages 39--45, 2003.

\bibitem{amershi2014power}
Saleema Amershi, Maya Cakmak, William~Bradley Knox, and Todd Kulesza.
\newblock Power to the people: The role of humans in interactive machine
  learning.
\newblock {\em Ai Magazine}, 35(4):105--120, 2014.

\bibitem{dudley2018review}
John~J Dudley and Per~Ola Kristensson.
\newblock A review of user interface design for interactive machine learning.
\newblock {\em ACM Transactions on Interactive Intelligent Systems (TiiS)},
  8(2):1--37, 2018.

\bibitem{fiebrink2011human}
Rebecca Fiebrink, Perry~R Cook, and Dan Trueman.
\newblock Human model evaluation in interactive supervised learning.
\newblock In {\em Proceedings of the SIGCHI Conference on Human Factors in
  Computing Systems}, pages 147--156, 2011.

\bibitem{arendt2019towards}
Dustin Arendt, Emily Saldanha, Ryan Wesslen, Svitlana Volkova, and Wenwen Dou.
\newblock Towards rapid interactive machine learning: Evaluating tradeoffs of
  classification without representation.
\newblock In {\em Proceedings of the International Conference on Intelligent
  User Interfaces}, page 591–602, 2019.

\bibitem{kulesza2015principles}
Todd Kulesza, Margaret Burnett, Weng-Keen Wong, and Simone Stumpf.
\newblock Principles of explanatory debugging to personalize interactive
  machine learning.
\newblock In {\em Proceedings of the International Conference on Intelligent
  User Interfaces}, pages 126--137, 2015.

\bibitem{talbot2009ensemblematrix}
Justin Talbot, Bongshin Lee, Ashish Kapoor, and Desney~S Tan.
\newblock Ensemblematrix: interactive visualization to support machine learning
  with multiple classifiers.
\newblock In {\em Proceedings of the SIGCHI Conference on Human Factors in
  Computing Systems}, pages 1283--1292, 2009.

\bibitem{kapoor2010interactive}
Ashish Kapoor, Bongshin Lee, Desney Tan, and Eric Horvitz.
\newblock Interactive optimization for steering machine classification.
\newblock In {\em Proceedings of the SIGCHI Conference on Human Factors in
  Computing Systems}, pages 1343--1352, 2010.

\bibitem{nakao2020use}
Yuri Nakao and Yusuke Sugano.
\newblock Use of machine learning by non-expert dhh people: Technological
  understanding and sound perception.
\newblock In {\em Proceedings of the Nordic Conference on Human-Computer
  Interaction: Shaping Experiences, Shaping Society}, pages 1--12, 2020.

\bibitem{zou2019object}
Zhengxia Zou, Zhenwei Shi, Yuhong Guo, and Jieping Ye.
\newblock Object detection in 20 years: A survey.
\newblock {\em arXiv preprint arXiv:1905.05055}, 2019.

\bibitem{zhao2019object}
Zhong-Qiu Zhao, Peng Zheng, Shou-tao Xu, and Xindong Wu.
\newblock Object detection with deep learning: A review.
\newblock {\em IEEE transactions on neural networks and learning systems},
  30(11):3212--3232, 2019.

\bibitem{papageorgiou1998general}
Constantine~P Papageorgiou, Michael Oren, and Tomaso Poggio.
\newblock A general framework for object detection.
\newblock In {\em Sixth International Conference on Computer Vision (IEEE Cat.
  No. 98CH36271)}, pages 555--562. IEEE, 1998.

\bibitem{long2015fully}
Jonathan Long, Evan Shelhamer, and Trevor Darrell.
\newblock Fully convolutional networks for semantic segmentation.
\newblock In {\em Proceedings of the IEEE conference on computer vision and
  pattern recognition}, pages 3431--3440, 2015.

\bibitem{yu2018methods}
Hongshan Yu, Zhengeng Yang, Lei Tan, Yaonan Wang, Wei Sun, Mingui Sun, and
  Yandong Tang.
\newblock Methods and datasets on semantic segmentation: A review.
\newblock {\em Neurocomputing}, 304:82--103, 2018.

\bibitem{taghanaki2021deep}
Saeid~Asgari Taghanaki, Kumar Abhishek, Joseph~Paul Cohen, Julien Cohen-Adad,
  and Ghassan Hamarneh.
\newblock Deep semantic segmentation of natural and medical images: a review.
\newblock {\em Artificial Intelligence Review}, 54(1):137--178, 2021.

\bibitem{radford2019language}
Alec Radford, Jeffrey Wu, Rewon Child, David Luan, Dario Amodei, Ilya
  Sutskever, et~al.
\newblock Language models are unsupervised multitask learners.
\newblock {\em OpenAI blog}, 1(8):9, 2019.

\bibitem{brown2020language}
Tom Brown, Benjamin Mann, Nick Ryder, Melanie Subbiah, Jared~D Kaplan, Prafulla
  Dhariwal, Arvind Neelakantan, Pranav Shyam, Girish Sastry, Amanda Askell,
  et~al.
\newblock Language models are few-shot learners.
\newblock {\em Advances in Neural Information Processing Systems},
  33:1877--1901, 2020.

\bibitem{gan2022vision}
Zhe Gan, Linjie Li, Chunyuan Li, Lijuan Wang, Zicheng Liu, Jianfeng Gao, et~al.
\newblock Vision-language pre-training: Basics, recent advances, and future
  trends.
\newblock {\em Foundations and Trends{\textregistered} in Computer Graphics and
  Vision}, 14(3--4):163--352, 2022.

\bibitem{lu2019vilbert}
Jiasen Lu, Dhruv Batra, Devi Parikh, and Stefan Lee.
\newblock Vilbert: Pretraining task-agnostic visiolinguistic representations
  for vision-and-language tasks.
\newblock {\em arXiv preprint arXiv:1908.02265}, 2019.

\bibitem{li2022blip}
Junnan Li, Dongxu Li, Caiming Xiong, and Steven Hoi.
\newblock Blip: Bootstrapping language-image pre-training for unified
  vision-language understanding and generation.
\newblock In {\em International Conference on Machine Learning}, pages
  12888--12900. PMLR, 2022.

\bibitem{hossain2019comprehensive}
MD~Zakir Hossain, Ferdous Sohel, Mohd~Fairuz Shiratuddin, and Hamid Laga.
\newblock A comprehensive survey of deep learning for image captioning.
\newblock {\em ACM Computing Surveys (CsUR)}, 51(6):1--36, 2019.

\bibitem{vinyals2015show}
Oriol Vinyals, Alexander Toshev, Samy Bengio, and Dumitru Erhan.
\newblock Show and tell: A neural image caption generator.
\newblock In {\em Proceedings of the IEEE conference on Computer Vision and
  Pattern Recognition}, pages 3156--3164, 2015.

\bibitem{pan2004automatic}
Jia-Yu Pan, Hyung-Jeong Yang, Pinar Duygulu, and Christos Faloutsos.
\newblock Automatic image captioning.
\newblock In {\em 2004 IEEE International Conference on Multimedia and Expo
  (ICME)(IEEE Cat. No. 04TH8763)}, volume~3, pages 1987--1990. IEEE, 2004.

\bibitem{ware2001interactive}
Malcolm Ware, Eibe Frank, Geoffrey Holmes, Mark Hall, and Ian~H Witten.
\newblock Interactive machine learning: letting users build classifiers.
\newblock {\em International Journal of Human-Computer Studies},
  55(3):281--292, 2001.

\bibitem{carney2020teachable}
Michelle Carney, Barron Webster, Irene Alvarado, Kyle Phillips, Noura Howell,
  Jordan Griffith, Jonas Jongejan, Amit Pitaru, and Alexander Chen.
\newblock Teachable machine: Approachable web-based tool for exploring machine
  learning classification.
\newblock In {\em Extended Abstracts of the CHI Conference on Human Factors in
  Computing Systems}, pages 1--8, 2020.

\bibitem{tatsuya2020investigating}
Tatsuya Ishibashi, Yuri Nakao, and Yusuke Sugano.
\newblock Investigating audio data visualization for interactive sound
  recognition.
\newblock In {\em Proceedings of the 25th International Conference on
  Intelligent User Interfaces}, pages 67--77, 2020.

\bibitem{kacorri2017people}
Hernisa Kacorri, Kris~M. Kitani, Jeffrey~P. Bigham, and Chieko Asakawa.
\newblock People with visual impairment training personal object recognizers:
  Feasibility and challenges.
\newblock In {\em Proceedings of the CHI Conference on Human Factors in
  Computing Systems}, page 5839–5849, 2017.

\bibitem{ahmetovic2020recog}
Dragan Ahmetovic, Daisuke Sato, Uran Oh, Tatsuya Ishihara, Kris Kitani, and
  Chieko Asakawa.
\newblock Recog: Supporting blind people in recognizing personal objects.
\newblock In {\em Proceedings of the CHI Conference on Human Factors in
  Computing Systems}, page 1–12, 2020.

\bibitem{liu2022interactive}
Tianyi Liu and Yusuke Sugano.
\newblock Interactive machine learning on edge devices with user-in-the-loop
  sample recommendation.
\newblock {\em IEEE Access}, 10:107346--107360, 2022.

\bibitem{fogarty2008cueflik}
James Fogarty, Desney Tan, Ashish Kapoor, and Simon Winder.
\newblock Cueflik: interactive concept learning in image search.
\newblock In {\em Proceedings of the SIGCHI Conference on Human Factors in
  Computing Systems}, pages 29--38, 2008.

\bibitem{pirrung2018sharkzor}
Meg Pirrung, Nathan Hilliard, Nancy O'Brien, Artem Yankov, Court~D Corley, and
  Nathan~O Hodas.
\newblock Sharkzor: Human in the loop ml for user-defined image classification.
\newblock In {\em Proceedings of the International Conference on Intelligent
  User Interfaces Companion}, pages 1--2, 2018.

\bibitem{hodas2016adding}
Nathan~Oken Hodas and Alex Endert.
\newblock Adding semantic information into data models by learning domain
  expertise from user interaction.
\newblock {\em arXiv preprint arXiv:1604.02935}, abs/1604.02935, 2016.

\bibitem{amershi2009overview}
Saleema Amershi, James Fogarty, Ashish Kapoor, and Desney Tan.
\newblock Overview based example selection in end user interactive concept
  learning.
\newblock In {\em Proceedings of the Annual ACM Symposium on User Interface
  Software and Technology}, pages 247--256, 2009.

\bibitem{kulesza2009fixing}
Todd Kulesza, Weng-Keen Wong, Simone Stumpf, Stephen Perona, Rachel White,
  Margaret~M Burnett, Ian Oberst, and Andrew~J Ko.
\newblock Fixing the program my computer learned: Barriers for end users,
  challenges for the machine.
\newblock In {\em Proceedings of the International Conference on Intelligent
  User Interfaces}, pages 187--196, 2009.

\bibitem{patel2008investigating}
Kayur Patel, James Fogarty, James~A Landay, and Beverly Harrison.
\newblock Investigating statistical machine learning as a tool for software
  development.
\newblock In {\em Proceedings of the SIGCHI Conference on Human Factors in
  Computing Systems}, pages 667--676, 2008.

\bibitem{li2011composing}
Siming Li, Girish Kulkarni, Tamara Berg, Alexander Berg, and Yejin Choi.
\newblock Composing simple image descriptions using web-scale n-grams.
\newblock In {\em Proceedings of the Conference on Computational Natural
  Language Learning}, pages 220--228, 2011.

\bibitem{aneja2018convolutional}
Jyoti Aneja, Aditya Deshpande, and Alexander~G Schwing.
\newblock Convolutional image captioning.
\newblock In {\em Proceedings of the IEEE conference on Computer Vision and
  Pattern Recognition}, pages 5561--5570, 2018.

\bibitem{you2016image}
Quanzeng You, Hailin Jin, Zhaowen Wang, Chen Fang, and Jiebo Luo.
\newblock Image captioning with semantic attention.
\newblock In {\em Proceedings of the IEEE conference on Computer Vision and
  Pattern Recognition}, pages 4651--4659, 2016.

\bibitem{cornia2020meshed}
Marcella Cornia, Matteo Stefanini, Lorenzo Baraldi, and Rita Cucchiara.
\newblock Meshed-memory transformer for image captioning.
\newblock In {\em Proceedings of the IEEE Conference on Computer Vision and
  Pattern Recognition}, pages 10578--10587, 2020.

\bibitem{liu2020interactive}
Junhao Liu, Kai Wang, Chunpu Xu, Zhou Zhao, Ruifeng Xu, Ying Shen, and Min
  Yang.
\newblock Interactive dual generative adversarial networks for image
  captioning.
\newblock In {\em Proceedings of the AAAI Conference on Artificial
  Intelligence}, volume~34, pages 11588--11595, 2020.

\bibitem{cornia2019show}
Marcella Cornia, Lorenzo Baraldi, and Rita Cucchiara.
\newblock Show, control and tell: A framework for generating controllable and
  grounded captions.
\newblock In {\em Proceedings of the IEEE Conference on Computer Vision and
  Pattern Recognition}, pages 8307--8316, 2019.

\bibitem{jia2020icap}
Zhengxiong Jia and Xirong Li.
\newblock icap: Interactive image captioning with predictive text.
\newblock In {\em Proceedings of the International Conference on Multimedia
  Retrieval}, pages 428--435, 2020.

\bibitem{hossain2021text}
Md~Zakir Hossain, Ferdous Sohel, Mohd~Fairuz Shiratuddin, Hamid Laga, and
  Mohammed Bennamoun.
\newblock Text to image synthesis for improved image captioning.
\newblock {\em IEEE Access}, 9:64918--64928, 2021.

\bibitem{chen2017show}
Tseng-Hung Chen, Yuan-Hong Liao, Ching-Yao Chuang, Wan-Ting Hsu, Jianlong Fu,
  and Min Sun.
\newblock Show, adapt and tell: Adversarial training of cross-domain image
  captioner.
\newblock In {\em Proceedings of the IEEE International Conference on Computer
  Vision}, pages 521--530, 2017.

\bibitem{zhao2020cross}
Wentian Zhao, Xinxiao Wu, and Jiebo Luo.
\newblock Cross-domain image captioning via cross-modal retrieval and model
  adaptation.
\newblock {\em IEEE Transactions on Image Processing}, 30:1180--1192, 2020.

\bibitem{long2020cross}
Cuirong Long, Xiaoshan Yang, and Changsheng Xu.
\newblock Cross-domain personalized image captioning.
\newblock {\em Multimedia Tools and Applications}, 79(45):33333--33348, 2020.

\bibitem{yang2018multitask}
Min Yang, Wei Zhao, Wei Xu, Yabing Feng, Zhou Zhao, Xiaojun Chen, and Kai Lei.
\newblock Multitask learning for cross-domain image captioning.
\newblock {\em IEEE Transactions on Multimedia}, 21(4):1047--1061, 2018.

\bibitem{zhao2017dual}
Wei Zhao, Wei Xu, Min Yang, Jianbo Ye, Zhou Zhao, Yabing Feng, and Yu~Qiao.
\newblock Dual learning for cross-domain image captioning.
\newblock In {\em Proceedings of the 2017 ACM on Conference on Information and
  Knowledge Management}, pages 29--38, 2017.

\bibitem{chen2021visualgpt}
Jun Chen, Han Guo, Kai Yi, Boyang Li, and Mohamed Elhoseiny.
\newblock Visualgpt: Data-efficient adaptation of pretrained language models
  for image captioning.
\newblock {\em arXiv preprint arXiv:2102.10407}, 2021.

\bibitem{li2020oscar}
Xiujun Li, Xi~Yin, Chunyuan Li, Pengchuan Zhang, Xiaowei Hu, Lei Zhang, Lijuan
  Wang, Houdong Hu, Li~Dong, Furu Wei, et~al.
\newblock Oscar: Object-semantics aligned pre-training for vision-language
  tasks.
\newblock In {\em Proceedings of the IEEE Conference on European Conference on
  Computer Vision}, pages 121--137. Springer, 2020.

\bibitem{zhou2022learning}
Kaiyang Zhou, Jingkang Yang, Chen~Change Loy, and Ziwei Liu.
\newblock Learning to prompt for vision-language models.
\newblock {\em International Journal of Computer Vision}, 130(9):2337--2348,
  2022.

\bibitem{zhou2022conditional}
Kaiyang Zhou, Jingkang Yang, Chen~Change Loy, and Ziwei Liu.
\newblock Conditional prompt learning for vision-language models.
\newblock In {\em Proceedings of the IEEE/CVF Conference on Computer Vision and
  Pattern Recognition}, pages 16816--16825, 2022.

\bibitem{desai2021virtex}
Karan Desai and Justin Johnson.
\newblock Virtex: Learning visual representations from textual annotations.
\newblock In {\em Proceedings of the IEEE Conference on Computer Vision and
  Pattern Recognition}, pages 11162--11173, 2021.

\bibitem{he2016deep}
Kaiming He, Xiangyu Zhang, Shaoqing Ren, and Jian Sun.
\newblock Deep residual learning for image recognition.
\newblock In {\em Proceedings of the IEEE conference on Computer Vision and
  Pattern Recognition}, pages 770--778, 2016.

\bibitem{vaswani2017attention}
Ashish Vaswani, Noam Shazeer, Niki Parmar, Jakob Uszkoreit, Llion Jones,
  Aidan~N Gomez, {\L}ukasz Kaiser, and Illia Polosukhin.
\newblock Attention is all you need.
\newblock In {\em Advances in Neural Information Processing Systems}, pages
  5998--6008, 2017.

\bibitem{deng2009imagenet}
Jia Deng, Wei Dong, Richard Socher, Li-Jia Li, Kai Li, and Li~Fei-Fei.
\newblock Imagenet: A large-scale hierarchical image database.
\newblock In {\em Proceedings of the IEEE conference on Computer Vision and
  Pattern Recognition}, pages 248--255. Ieee, 2009.

\bibitem{lin2014microsoft}
Tsung-Yi Lin, Michael Maire, Serge Belongie, James Hays, Pietro Perona, Deva
  Ramanan, Piotr Doll{\'a}r, and C~Lawrence Zitnick.
\newblock Microsoft coco: Common objects in context.
\newblock In {\em Proceedings of the IEEE conference on European Conference on
  Computer Vision}, pages 740--755. Springer, 2014.

\bibitem{loshchilov2017decoupled}
Ilya Loshchilov and Frank Hutter.
\newblock Decoupled weight decay regularization.
\newblock {\em arXiv preprint arXiv:1711.05101}, 2017.

\bibitem{wu2014learning}
Qi~Wu, Hongping Cai, and Peter Hall.
\newblock Learning graphs to model visual objects across different depictive
  styles.
\newblock In {\em Proceedings of the IEEE conference on European Conference on
  Computer Vision}, pages 313--328, 2014.

\bibitem{yao2011human}
Bangpeng Yao, Xiaoye Jiang, Aditya Khosla, Andy~Lai Lin, Leonidas Guibas, and
  Li~Fei-Fei.
\newblock Human action recognition by learning bases of action attributes and
  parts.
\newblock In {\em Proceedings of the IEEE conference on International
  Conference on Computer Vision}, pages 1331--1338, 2011.

\bibitem{feifei2004learning}
Li~Fei-Fei, Rob Fergus, and Pietro Perona.
\newblock Learning generative visual models from few training examples: An
  incremental bayesian approach tested on 101 object categories.
\newblock In {\em Conference on Computer Vision and Pattern Recognition
  Workshop}, pages 178--178, 2004.

\bibitem{liu2016deepfashion}
Ziwei Liu, Ping Luo, Shi Qiu, Xiaogang Wang, and Xiaoou Tang.
\newblock Deepfashion: Powering robust clothes recognition and retrieval with
  rich annotations.
\newblock In {\em Proceedings of IEEE Conference on Computer Vision and Pattern
  Recognition}, June 2016.

\bibitem{bossard2014food}
Lukas Bossard, Matthieu Guillaumin, and Luc Van~Gool.
\newblock Food-101--mining discriminative components with random forests.
\newblock In {\em Proceedings of the IEEE conference on European Conference on
  Computer Vision}, pages 446--461, 2014.

\bibitem{shao2018crowdhuman}
Shuai Shao, Zijian Zhao, Boxun Li, Tete Xiao, Gang Yu, Xiangyu Zhang, and Jian
  Sun.
\newblock Crowdhuman: A benchmark for detecting human in a crowd.
\newblock {\em arXiv preprint arXiv:1805.00123}, 2018.

\bibitem{zhang2017age}
Zhifei Zhang, Yang Song, and Hairong Qi.
\newblock Age progression/regression by conditional adversarial autoencoder.
\newblock In {\em Proceedings of the IEEE conference on Computer Vision and
  Pattern Recognition}. IEEE, 2017.

\bibitem{gould2009decomposing}
Stephen Gould, Richard Fulton, and Daphne Koller.
\newblock Decomposing a scene into geometric and semantically consistent
  regions.
\newblock In {\em Proceedings of the IEEE conference on International
  Conference on Computer Vision}, pages 1--8, 2009.

\bibitem{zhu2016face}
Xiangyu Zhu, Zhen Lei, Xiaoming Liu, Hailin Shi, and Stan~Z Li.
\newblock Face alignment across large poses: A 3d solution.
\newblock In {\em Proceedings of the IEEE conference on Computer Vision and
  Pattern Recognition}, pages 146--155, 2016.

\bibitem{spithourakis2018numeracy}
Georgios Spithourakis and Sebastian Riedel.
\newblock Numeracy for language models: Evaluating and improving their ability
  to predict numbers.
\newblock In {\em Proceedings of the 56th Annual Meeting of the Association for
  Computational Linguistics (Volume 1: Long Papers)}, pages 2104--2115, 2018.

\bibitem{thawani2021numeracy}
Avijit Thawani, Jay Pujara, and Filip Ilievski.
\newblock Numeracy enhances the literacy of language models.
\newblock In {\em Proceedings of the 2021 conference on empirical methods in
  natural language processing}, pages 6960--6967, 2021.

\bibitem{hart1988development}
Sandra~G Hart and Lowell~E Staveland.
\newblock Development of nasa-tlx (task load index): Results of empirical and
  theoretical research.
\newblock In {\em Advances in Psychology}, volume~52, pages 139--183. Elsevier,
  1988.

\end{thebibliography}

\end{document}